\documentclass[aps,pre,twocolumn,superscriptaddress,notitlepage]{revtex4-2}
\usepackage{amsmath}
\usepackage{amsfonts}
\usepackage{amssymb}
\usepackage{lmodern,dsfont}
\usepackage{graphicx}
\usepackage[usenames,dvipsnames]{xcolor}
\usepackage{bm}
\usepackage{xr}
\usepackage[english=american]{csquotes}

\newcommand{\av}[1]{\langle #1 \rangle}
\newcommand{\Av}[1]{\left\langle #1 \right\rangle}
\newcommand{\nn}{\nonumber \\}
\newcommand{\n}{\nonumber}

\newcommand{\grad}{\bm{\nabla}}

\renewcommand{\eqref}[1]{Eq.~(\ref{#1})}

\begin{document}

\author{Andreas Dechant}
\affiliation{Department of Physics \#1, Graduate School of Science, Kyoto University, Kyoto 606-8502, Japan}
\author{J{\'e}r{\^o}me Garnier-Brun}
\affiliation{Chair of Econophysics and Complex Systems, \'Ecole polytechnique, 91128 Palaiseau Cedex, France}
\affiliation{LadHyX, CNRS, École polytechnique, Institut Polytechnique de Paris, 91120 Palaiseau, France}
\author{Shin-ichi Sasa}
\affiliation{Department of Physics \#1, Graduate School of Science, Kyoto University, Kyoto 606-8502, Japan}
\title{Thermodynamic bounds on correlation times}
\date{\today}

\begin{abstract}
We derive a variational expression for the correlation time of physical observables in steady-state diffusive systems.
As a consequence of this variational expression, we obtain lower bounds on the correlation time, which provide speed limits on the self-averaging of observables.
In equilibrium, the bound takes the form of a tradeoff relation between the long- and short-time fluctuations of an observable.
Out of equilibrium, the tradeoff can be violated, leading to an acceleration of self-averaging. 
We relate this violation to the steady-state entropy production rate, as well as the geometric structure of the irreversible currents, giving rise to two complementary speed limits.
One of these can be formulated as a lower estimate on the entropy production from the measurement of time-symmetric observables.
Using an illustrating example, we show the intricate behavior of the correlation time out of equilibrium for different classes of observables and how this can be used to partially infer dissipation even if no time-reversal symmetry breaking can be observed in the trajectories of the observable.
\end{abstract}

\maketitle

A characteristic property of noisy systems is that, even in a steady state, where the ensemble probability does not change, there are still dynamics in the system.
Characterizing the instantaneous configuration of the system at time $t$ by a collection of degrees of freedom $\bm{x}(t) = (x_1(t),\ldots,x_d(t))$, we denote by $p_\text{st}(\bm{x})$ the time-independent steady-state probability of observing a given configuration.
Because of the presence of noise, the instantaneous configuration $\bm{x}(t)$ exhibits time-dependent fluctuations. 
So does any configuration-dependent observable $z(\bm{x}(t))$, whereas the ensemble average $\av{z}_\text{st}$ is time-independent.
For ergodic systems, the connection between the fluctuating single realizations and the ensemble is that time averaged observables $\bar{z}_\tau = \int_0^\tau dt \ z(\bm{x}(t))/\tau$ converge to their ensemble averages in the long-time limit, $\lim_{\tau \rightarrow \infty} \bar{z}_\tau = \av{z}_\text{st}$.
For a sufficiently long measurement time $\tau$, this self-averaging allows us to deduce ensemble-averaged observables from a single realization.

For any finite time, however, $\bar{z}_\tau$ fluctuates, and we characterize these fluctuations by the variance $\text{Var}(\bar{z}_\tau)$.
Ergodicity then implies $\lim_{\tau \rightarrow \infty} \text{Var}(\bar{z}_\tau) = 0$.
We can characterize the speed of the self-averaging process by defining the correlation time $\tau^z$ of the observable as
\begin{align}
\tau^z = \frac{\int_0^\infty dt \ \text{Cov}(z(t),z(0)}{\text{Var}_\text{st}(z)} \label{green-kubo},
\end{align}
where $\text{Cov}$ denotes the covariance and $\text{Var}_\text{st}(z)$ the variance of $z(\bm{x})$ in the steady state.
Intuitively, $\tau^z$ measures the typical timescale on which the correlations of $z(\bm{x}(t))$ decay.
For time-lags $\tau$ longer than $\tau^z$, $z(\bm{x}(t+\tau))$ is approximately independent of $z(\bm{x}(t))$, and the time-average can be regarded as a sum of independent random variables. 
We then have 
\begin{align}
\frac{\text{Var}(\bar{z}_\tau)}{2 \text{Var}_\text{st}(z)} \simeq \frac{\tau^z}{\tau},
\end{align}
in agreement with the central limit theorem.

While the \eqref{green-kubo} gives a prescription to compute $\tau^z$, an explicit expression is only available in simple cases \cite{Dec11} and the relation to other physical quantities is not readily apparent.
In this work, we derive lower bounds on $\tau^z$ in and out of equilibrium, which constitute speed limits on the self-averaging of observables.
In contrast to existing speed limits \cite{Oku18,Shi18,Vo20,Fal20,Nic20,Ito20,Van21,Yos21}, which describe the transition between different ensemble states in stochastic systems, our speed limits characterize the decay of correlations in a steady state as a consequence of the noisy dynamics.
In particular, they highlight the influence of irreversible currents and their geometric structure on correlations and self-averaging out of equilibrium.

\textit{Physical setup.} 
For the sake of concreteness, we will focus on the overdamped Langevin dynamics ($\bm{x} \in \mathbb{R}^d$)
\begin{align}
\dot{\bm{x}}(t) = \bm{a}(\bm{x}(t)) + \bm{G} \bm{\xi}(t) \label{langevin},
\end{align}
with drift vector $\bm{a}(\bm{x})$ and full rank matrix $\bm{G}$, where $\bm{\xi}(t)$ is a vector of mutually independent Gaussian white noises.
We stress that similar bounds can be derived for jump processes; we will address this case in a forthcoming publication.
If the drift vector satisfies the potential condition $\bm{a}(\bm{x}) = \bm{B} \grad \phi(\bm{x})$, where $\bm{B} = \bm{G} \bm{G}^\text{T}$ is the positive definite diffusion matrix, then the steady-state of \eqref{langevin} is the Boltzmann-Gibbs equilibrium $p_\text{st}(\bm{x}) = p_\text{eq}(\bm{x}) = e^{\phi(\bm{x})}/\int d\bm{x} \ e^{\phi(\bm{x})}$ and the system satisfies detailed balance \cite{Ris86}.
For generic $\bm{a}(\bm{x})$, however, the steady state is out of equilibrium and exhibits a non-vanishing rate of entropy production
\begin{align}
\sigma_\text{st} = \av{\sigma}_\text{st} \quad \text{with} \quad \sigma(\bm{x}) = \bm{\nu}_\text{st}(\bm{x}) \cdot \bm{B}^{-1} \bm{\nu}_\text{st}(\bm{x}) \label{meanvel-sigma} .
\end{align}
Here $\bm{\nu}_\text{st}(\bm{x}) = \bm{a}(\bm{x}) - \bm{B} \grad \ln p_\text{st}(\bm{x})$ is called the local mean velocity; it describes the irreversible currents in the system as a consequence of broken detailed balance.

\textit{Main results.}
For an equilibrium system, we find the variational expression and lower bound 
\begin{align}
\tau^z_\text{eq} = \sup_{\chi} \Bigg[\frac{\frac{\text{Cov}_\text{eq}(z,\chi)^2}{\text{Var}_\text{eq}(\chi)}}{D^{\chi}} \Bigg] \geq \frac{\text{Var}_\text{eq}(z)}{D^z} \label{speed-limit-eq},
\end{align}
where $D^z$ quantifies the short-time fluctuations of the displacement $dz = z(\bm{x}(t+dt)) - z(\bm{x}(t))$,
\begin{align}
D^z = \lim_{dt \rightarrow 0} \Bigg[ \frac{\text{Var}(dz)}{2 dt} \Bigg] = \av{\grad z \cdot \bm{B} \grad z}_\text{eq} \label{short-time-diffusion} .
\end{align}
Since $\tau^z$ governs the long-time fluctuations of the time-average, \eqref{speed-limit-eq} implies a tradeoff between short- and long-time fluctuations of the observable.
The short-time fluctuations reflect the reversible diffusive motion in the system, which is the only way in which an equilibrium system can explore its configuration space and thereby self-average.
We note that the same type of relation also applies to underdamped Langevin dynamics (see section \ref{sec-underdamped} of the appendix) and jump processes.
Therefore, \eqref{correlation-intrinsic-variational} constitutes a universal tradeoff between diffusion and self-averaging that applies to a wide range of equilibrium processes.
Rather than a particular observable, we can also characterize the self-averaging behavior of the system by defining the intrinsic correlation time
\begin{align}
\tau^* = \sup_z \big[ \tau^z \big] \label{correlation-time-intrinsic},
\end{align}
that is, by considering the observable with the slowest self-averaging speed.
For the latter, we find the identity
\begin{align}
\tau^*_\text{eq} = \sup_{\chi} \Bigg[\frac{\text{Var}_\text{eq}(\chi)}{D^\chi} \Bigg] = \frac{1}{\lambda_\text{eq}^1} \label{intrinsic-eigenvalue-eq},
\end{align}
and thus the bound \eqref{speed-limit-eq} becomes tight for the slowest observable.
Crucially, this expression is equivalent to a well-know variational formula (see e.~g.~chapter 6.6.2 in Ref.~\cite{Ris86}) for the first non-zero eigenvalue of the generator of the dynamics, $\lambda_\text{eq}^1$.
Since the eigenvalue governs the asymptotic approach of the system towards equilibrium, $|p_t(\bm{x}) - p_\text{eq}(\bm{x})| \sim e^{-\lambda_\text{eq}^1 t}$, this relation formally establishes that both correlations in equilibrium and the relaxation towards equilibrium are governed by the same timescale. 

For a nonequilibrium system, on the other hand, the correlation time is reduced, $\tau^z \leq \tau^z_\text{eq}$, corresponding to faster self-averaging \cite{Hwa15,Dun16,Dun17,Cog21}. 
Here $\tau^z_\text{eq}$ is the correlation time in the equilibrium system with the same steady state. 
Intuitively, nonequilibrium systems can explore their configuration space not only by reversible diffusive motion, but also by irreversible, directed motion in the form of currents, which provide another mechanism for self-averaging.
The thermodynamic consequence of the irreversible currents is dissipation, characterized by the entropy production rate $\sigma_\text{st} > 0$.
Indeed, an appropriate nonequilibrium generalization of \eqref{speed-limit-eq} can be expressed in terms of entropy production,
\begin{align}
\tau^z \geq \frac{\text{Var}_\text{st}(z)}{D^z + \sigma_\text{st} \text{Var}_\sigma(z)} \label{speed-limit-sigma},
\end{align}
where $\text{Var}_\sigma(z)$ is the variance of $z(\bm{x})$ with respect to a probability distribution reweighted according to the local rate of entropy production.
While the dynamics of the system is accelerated by driving it out of equilibrium, \eqref{speed-limit-sigma} states that there is a minimal amount of dissipation associated with this acceleration, and we refer to it as the dissipation speed limit.

Moreover, we derive a complementary lower bound on the correlation time,
\begin{align}
\tau^z \geq \sup_{\chi_{\perp}} \Bigg[\frac{\frac{\text{Cov}_\text{st}(z,\chi_\perp)^2}{\text{Var}_\text{st}(\chi_{\perp})}}{D^{\chi_{\perp}}} \Bigg] \label{speed-limit-geometric} .
\end{align}
While this resembles the equilibrium result \eqref{speed-limit-eq}, the maximum is restricted to functions whose gradient is orthogonal to the irreversible currents $\grad \chi_{\perp}(\bm{x}) \cdot \bm{\nu}_\text{st}(\bm{x})$, and thus the right-hand side is always smaller than the equilibrium value.
Intuitively, if the irreversible currents in the system only flow along level lines of an observable, then they do not contribute to changes in its value and therefore do not accelerate the self-averaging.
In contrast to \eqref{speed-limit-sigma}, this bound is purely geometric as it does not depend on the magnitude of the currents; we refer to it as the geometric speed limit.
In particular, this implies that the acceleration of the self-averaging tends to saturate in the strong driving limit.

\textit{Variational formula and equilibrium speed limits.}
We now outline how the different speed limits may be derived from a variational formula for the correlation time,
\begin{align}
\tau^z = \sup_{\chi} \Bigg[ \frac{\frac{\text{Cov}_\text{st}(z,\chi)^2}{\text{Var}_\text{st}(z)}}{\av{\grad \chi \cdot \bm{B} \grad \chi}_\text{st} + \sup_{\eta} \Big[ \frac{\av{\chi \grad \eta \cdot \bm{\nu}_\text{st}}_\text{st}^2}{\av{\grad \eta \cdot \bm{B} \grad \eta}_\text{st}} \Big]} \Bigg] \label{correlation-variational} ,
\end{align}
where the maxima are taken with respect to differentiable functions $\chi(\bm{x})$ and $\eta(\bm{x})$.
This replaces the problem of determining the transition probability with solving a variational problem in terms of steady-state averages.
While an exact solution is equally hard to obtain, \eqref{correlation-variational} is immediately useful for deriving bounds, which are not apparent from \eqref{green-kubo}.
The derivation of \eqref{correlation-variational} is provided in section \ref{sec-variational} of the appendix.
The second term in the denominator is positive and vanishes in equilibrium, where $\bm{\nu}_\text{st}(\bm{x}) \equiv 0$.
Thus, we immediately conclude that $\tau^z \leq \tau^z_\text{eq}$, the latter being the correlation time in the (unique) equilibrium system with the same steady state $p_\text{eq}(\bm{x}) = p_\text{st}(\bm{x})$ and diffusion matrix $\bm{B}$,
\begin{align}
\tau^z_\text{eq} = \sup_{\chi} \Bigg[ \frac{\frac{\text{Cov}_\text{eq}(z,\chi)^2}{\text{Var}_\text{st}(z)}}{\av{\grad \chi \cdot \bm{B} \grad \chi}_\text{eq}} \Bigg] \geq \frac{\text{Var}_\text{eq}(z)}{\av{\grad z \cdot \bm{B} \grad z}_\text{eq}} \label{correlation-variational-equilibrium}.
\end{align}
The inequality follows by choosing $\chi(\bm{x}) = z(\bm{x})$ and noting that $\text{Cov}_\text{eq}(z,z) = \text{Var}_\text{eq}(z)$.
Identifying the term in the denominator with $D^z$, \eqref{short-time-diffusion} (see section \ref{sec-displacement} of the appendix), yields \eqref{speed-limit-eq}.
Using \eqref{correlation-time-intrinsic} and taking the maximum over $z(\bm{x})$ in \eqref{correlation-variational}, we also obtain a variational formula for the intrinsic correlation time
\begin{align}
\tau^* =  \sup_{\chi} \Bigg[ \frac{\text{Var}_\text{st}(\chi)}{\av{\grad \chi \cdot \bm{B} \grad \chi}_\text{st} + \sup_{\eta} \Big[ \frac{\av{\chi \grad \eta \cdot \bm{\nu}_\text{st}}_\text{st}^2}{\av{\grad \eta \cdot \bm{B} \grad \eta}_\text{st}}\Big]}  \Bigg] \label{correlation-intrinsic-variational} ,
\end{align}
which immediately gives \eqref{intrinsic-eigenvalue-eq} in equilibrium.

\textit{Nonequilibrium speed limits.}
While, in principle, we can obtain a lower bound on $\tau^z$ out of equilibrium by any specific choice of $\chi(\bm{x})$ in \eqref{correlation-variational}, this still involves the maximization over $\eta(\bm{x})$ and is thus not explicit.
However, we can further bound the second term in the denominator of \eqref{correlation-variational}.
First, we note that we have
\begin{align}
\av{\chi \grad \eta \cdot \bm{\nu}_\text{st}}_\text{st}^2 &= \av{(\chi - \chi_0) \grad \eta \cdot \bm{\nu}_\text{st}}_\text{st}^2 \\
&\leq \av{(\chi-\chi_0)^2 \bm{\nu}_\text{st} \cdot \bm{B}^{-1} \bm{\nu}_\text{st}}_\text{st} \av{\grad \eta \cdot \bm{B} \grad \eta}_\text{st} \n .
\end{align}
In the first step, we used that, from the steady-state condition of the Fokker-Planck equation, the local mean velocity satisfies $\grad \cdot (\bm{\nu}_\text{st}(\bm{x}) p_\text{st}(\bm{x}) ) = 0$, such that $\av{\chi_0 \grad \eta \cdot \bm{\nu}_\text{st}}_\text{st} = 0$ for any constant $\chi_0$ after integrating by parts.
In the second step, we applied the Cauchy-Schwarz inequality.
The second factor precisely cancels the one in the denominator, so that we have the lower bound
\begin{align}
\tau^z \geq \sup_{\chi} \Bigg[ \frac{\frac{\text{Cov}_\text{st}(z,\chi)^2}{\text{Var}_\text{st}(z)}}{\av{\grad \chi \cdot \bm{B} \grad \chi}_\text{st} + \av{\chi^2 \sigma}_\text{st} - \frac{\av{\chi \sigma}_\text{st}^2}{\sigma_\text{st}}} \Bigg],
\end{align}
where we maximized with respect to $\chi_0$ and used the definition of the local entropy production rate \eqref{meanvel-sigma}.
Introducing the entropy-rescaled probability density,
\begin{align}
p_\sigma(\bm{x}) = \frac{\sigma(\bm{x})}{\sigma_\text{st}} p_\text{st}(\bm{x}), 
\end{align}
and recalling \eqref{short-time-diffusion}, this can be written as
\begin{align}
\tau^z \geq \sup_{\chi} \Bigg[ \frac{\frac{\text{Cov}_\text{st}(z,\chi)^2}{\text{Var}_\text{st}(z)}}{D^\chi + \sigma_\text{st} \text{Var}_\sigma(\chi)} \Bigg] \label{lower-bound-entropy-variance},
\end{align}
which yields \eqref{speed-limit-sigma} after choosing $\chi(\bm{x}) = z(\bm{x})$.

On the other hand, integrating by parts, we have 
\begin{align}
\av{\chi \grad \eta \cdot \bm{\nu}_\text{st}}_\text{st}^2 &= \av{\eta \grad \chi \cdot \bm{\nu}_\text{st}}_\text{st}^2 . 
\end{align}
For any $\chi_\perp(\bm{x})$ that satisfies $\grad \chi_\perp(\bm{x}) \cdot \bm{\nu}_\text{st}(\bm{x}) = 0$, this term vanishes and we obtain \eqref{speed-limit-geometric}.
Thus, whenever there exists a function $\chi(\bm{x})$ whose gradient is orthogonal to the currents and $\text{Cov}_\text{st}(z,\chi) \neq 0$, we obtain a nonzero lower bound whose value is independent of the magnitude of the currents, but only depends on their geometric structure.
If the observable itself satisfies $\grad z(\bm{x}) \cdot \bm{\nu}_\text{st}(\bm{x})$, we can choose $\chi_\perp(\bm{x}) = z(\bm{x})$ in \eqref{speed-limit-geometric} and have
\begin{align}
\tau^{z_\perp} \geq \frac{\text{Var}_\text{st}(z_\perp)}{D^{z_\perp}} \label{speed-limit-geometric-observable} .
\end{align}
This implies that observables, whose level lines are parallel to the irreversible currents, obey the equilibrium tradeoff between their short- and long-time fluctuations.

\textit{Estimation of entropy production.}
The fact that \eqref{speed-limit-sigma} relates the correlation time out of equilibrium to dissipation suggests that it may be possible to estimate the latter by measuring the correlation time.
To make this relation explicit, we note that, if $\chi_\text{min} \leq \chi(\bm{x}) \leq \chi_\text{max}$ is a bounded function with range $\Delta \chi = \chi_\text{max} - \chi_\text{min}$, Popoviciu's inequality yields an upper bound on the variance, $\text{Var}_\sigma(\chi) \leq \Delta \chi^2/4$.
Plugging this into \eqref{lower-bound-entropy-variance} and solving for $\sigma_\text{st}$ yields
\begin{align}
\sigma_\text{st} \geq \frac{4}{\Delta \chi^2} \bigg( \frac{2 \text{Cov}_\text{st}(\chi,z)^2}{\text{Var}(\bar{z}_\tau)} - D^\chi \bigg) \label{correlation-uncertainty-2}.
\end{align}
Thus, we can obtain a lower bound on the rate of entropy production by measuring the fluctuations of the time-average of an observable and its steady-state correlations with any bounded observable.
In particular, if $z(\bm{x})$ itself is bounded, then we have
\begin{align}
\sigma_\text{st} \geq \frac{4}{\Delta z^2} \bigg( \frac{2 \text{Var}_\text{st}(z)^2}{\text{Var}(\bar{z}_\tau)} - D^z \bigg) \label{correlation-uncertainty} ,
\end{align}
which allows us to estimate the entropy production by measuring how much the short- and long-time fluctuations violate the equilibrium tradeoff \eqref{speed-limit-eq}.
We remark that several bounds relating entropy production to measurable quantities have recently been obtained, most famously the thermodynamic uncertainty relation \cite{Bar15,Gin16} and its many generalizations \cite{Pie17,Mac18,Dec18c,Has19,Koy19,Koy20,Liu20}.
However, most of these bounds rely on the measurement of some time-antisymmetric observable like a time-integrated current and its fluctuations.
A notable exception is Ref.~\cite{Koy20}, where a bound in terms of time-symmetric observables was obtained, which, however, vanishes in the steady state and only estimate the excess part of the entropy production.
To our knowledge, the lower bound \eqref{correlation-uncertainty} is the first result that only involves time-symmetric quantities in the steady state.

\textit{Illustration.}
\begin{figure*}
\includegraphics[width=.24\textwidth]{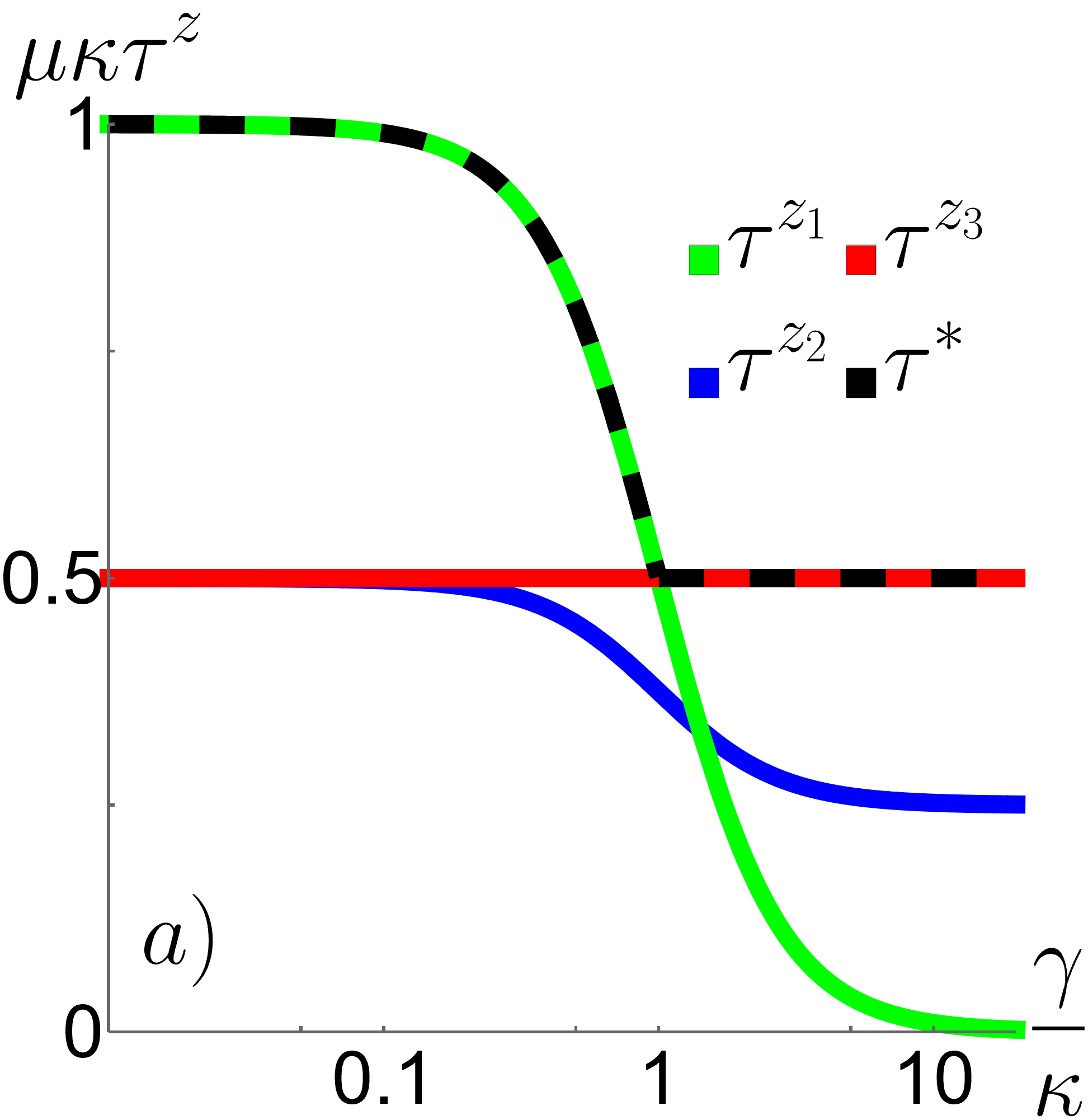}
\includegraphics[width=.24\textwidth]{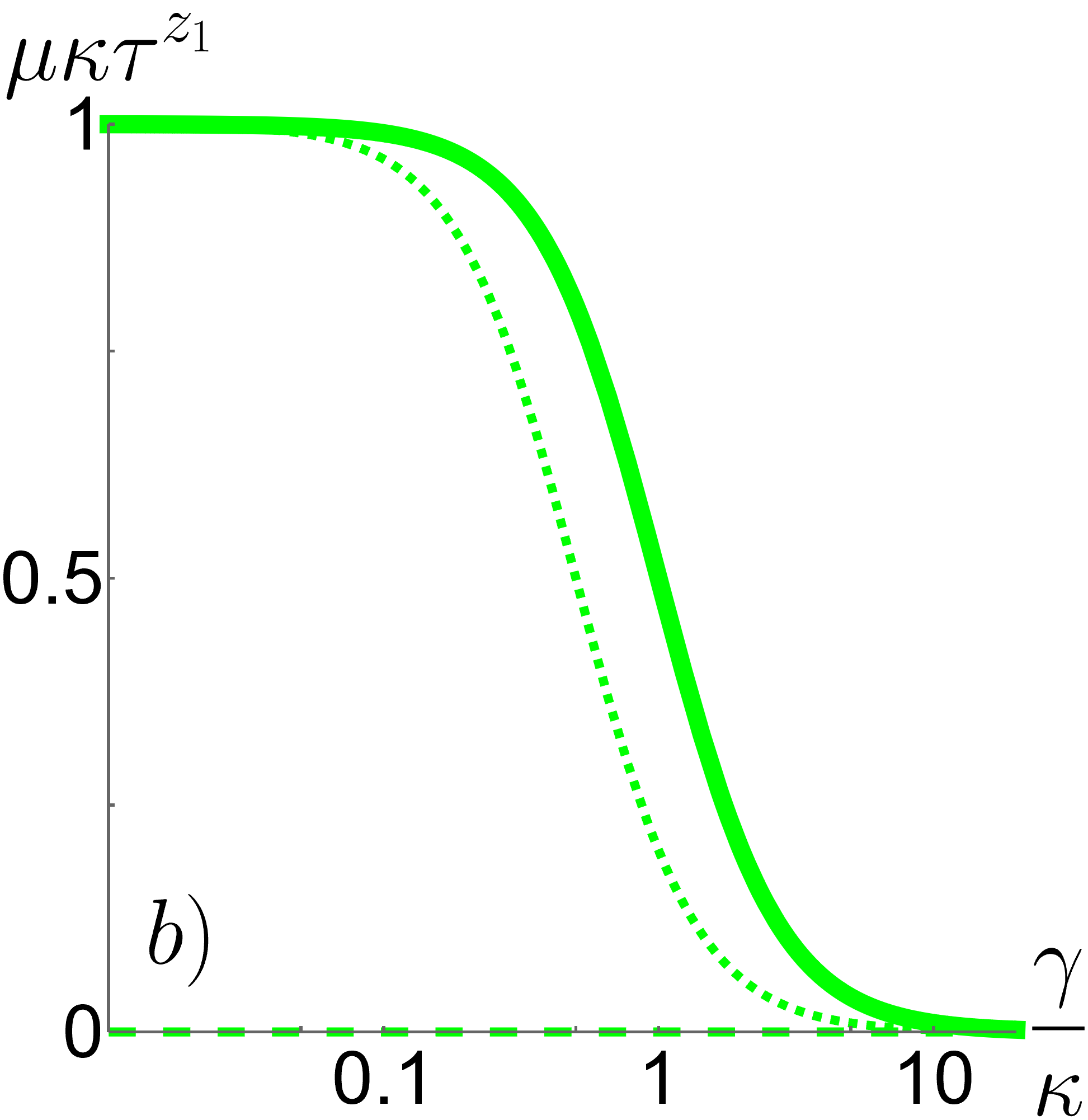}
\includegraphics[width=.24\textwidth]{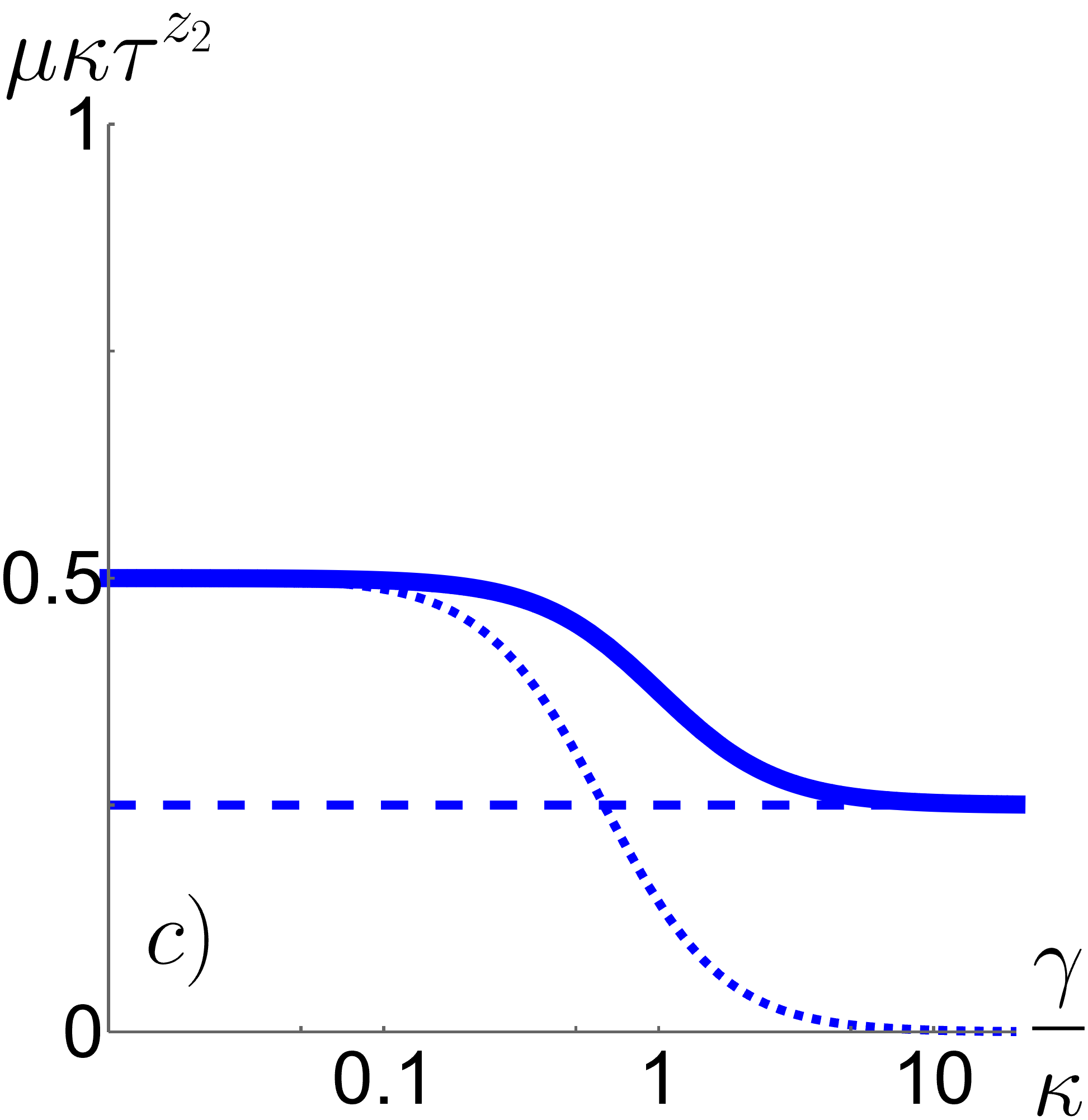}
\includegraphics[width=.24\textwidth]{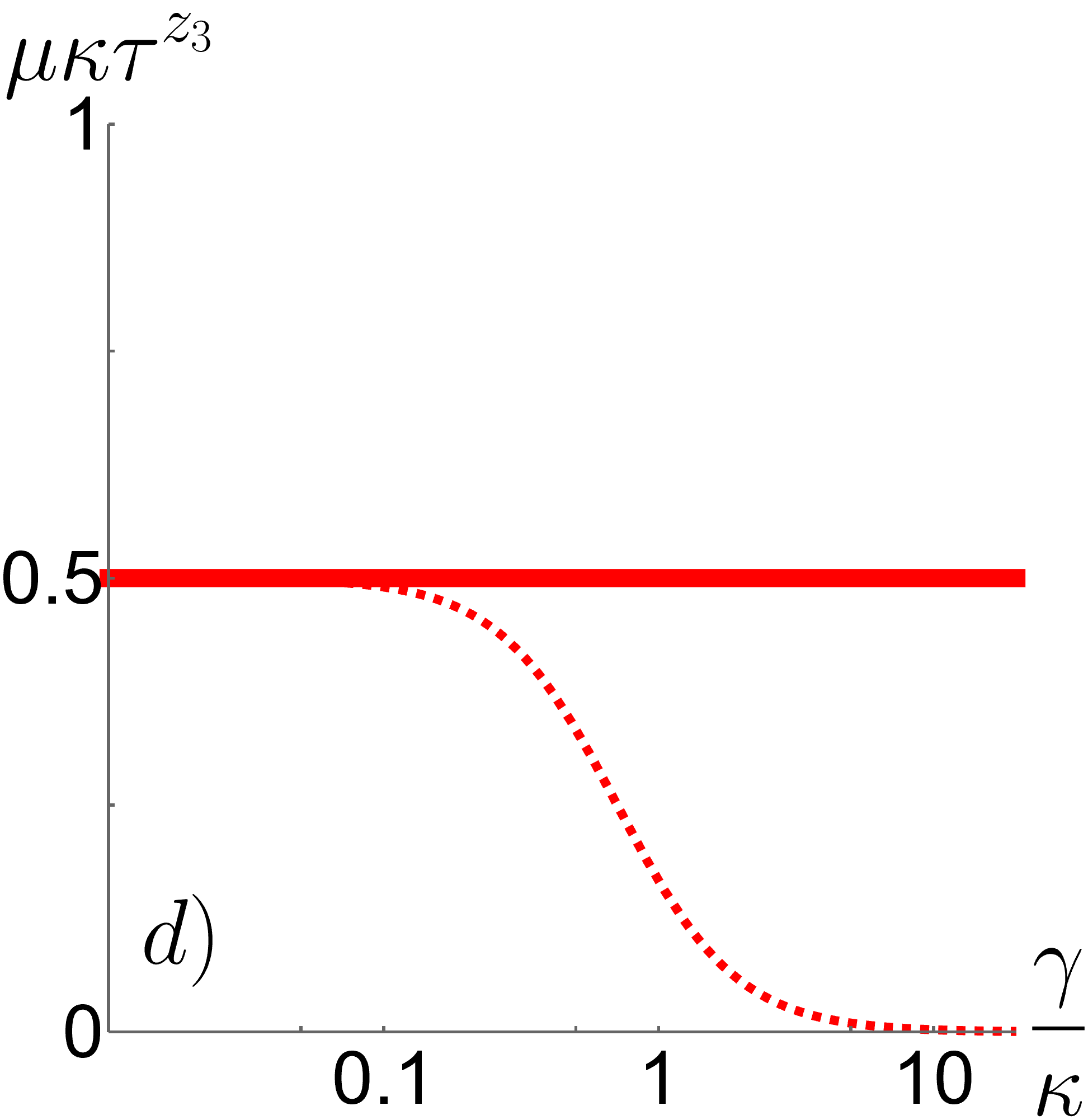}
\caption{The correlation times \eqref{parabolic-correlation} and corresponding speed limits \eqref{parabolic-speed-limit} for three observables of a driven Brownian particle, as a function of the driving strength.
a) The exact results for the correlation times of the observables $z_1 = x_1$, $z_2 = x_1^2$ and $z_3 = x_1^2 + x_2^2$ (colored lines), as well as the intrinsic (maximal) correlation time $\tau^*$ (black dashed line).
b)-d) The dissipative speed limit (\eqref{speed-limit-sigma}, dotted lines) and the geometric speed limit (\eqref{speed-limit-geometric}, dashed lines) on the correlation times compared with the exact results (solid lines).
Note that for $z_3$, the exact value and the geometric speed limit are identical.}
\label{fig-parabolic}
\end{figure*}
In order to illustrate the above speed limits for a concrete system, we consider a two-dimensional Brownian particle trapped in a parabolic potential $U(\bm{x}) = \kappa (x_1^2 + x_2^2)/2$ and driven by the nonconservative force $\bm{F}_\text{nc}(\bm{x}) = \gamma (x_2,-x_1)$.
The particle is in contact with a heat bath at temperature $T$ and its mobility is $\mu$.
The steady state of this system is given by the Gaussian $p_\text{st} = e^{-U(\bm{x})/T}/\int d\bm{x} \hspace{0.05cm} e^{-U(\bm{x})/T}$, independent of the driving strength $\gamma$, while the local mean velocity is given by $\bm{\nu}_\text{st}(\bm{x}) = \mu \bm{F}_\text{nc}(\bm{x})$.
Since the forces in this system are linear, we can compute the transition probability analytically, see section \ref{sec-parabolic-transition} of the appendix for the details of the calculation.
We consider the observables $z_1(\bm{x}) = x_1$, $z_2(\bm{x}) = x_1^2$ and $z_3(\bm{x}) = x_1^2 + x_2^2$; their correlation times are
\begin{align}
\tau^{z_1} = \frac{1}{\mu \kappa \big( 1 + \frac{\gamma^2}{\kappa^2} \big)}, \quad \tau^{z_2} = \frac{2 + \frac{\gamma^2}{\kappa^2}}{4 \mu \kappa\big(1 + \frac{\gamma^2}{\kappa^2} \big)}, \quad \tau^{z_3} = \frac{1}{2 \mu \kappa} \label{parabolic-correlation} ,
\end{align}
which are shown graphically in Fig.~\ref{fig-parabolic}a).
In equilibrium ($\gamma = 0$), all three observables satisfy the corresponding speed limit \eqref{speed-limit-eq} with equality and thus saturate the tradeoff between short- and long-time fluctuations.
Out of equilibrium, the observables exhibit a markedly different behavior:
For $z_1 = x_1$, the correlation time tends to zero and its self-averaging becomes arbitrarily fast with increasing driving strength.
For $z_2 = x_1^2$, the correlation time also decreases when driving the system out of equilibrium, however, it saturates in the limit of strong driving, indicating that stronger driving cannot speed up its self-averaging arbitrarily.
The correlation time of $z_3 = x_1^2 + x_2^2$ is not affected at all by the driving and this particular driving force cannot speed up its self-averaging.
By contrast, the intrinsic correlation time is (see section \ref{sec-parabolic-intrinsic} for the calculation)
\begin{align}
\tau^* = \max \big(\tau^{z_1},\tau^{z_3}\big) = \tau^*_\text{eq} \max \bigg( \frac{1}{1 + \frac{\gamma^2}{\kappa^2}}, \frac{1}{2} \bigg) ,
\end{align}
where $\tau_\text{eq}^* = 1/(\mu \kappa)$ is the equilibrium value.
Interestingly, the worst-case observable exhibiting the slowest self-averaging depends on the parameters of the system, which leads to a non-smooth behavior of the intrinsic correlation time.

We now turn to the nonequilibrium speed limits, denoting the the lower bound obtained from the dissipation speed limit \eqref{speed-limit-sigma} by $\tau^z_\text{diss}$ and the given by the geometric speed limit \eqref{speed-limit-geometric} by $\tau^z_\text{geom}$.
For the observables defined above, we have
\begin{subequations}
\begin{align}
\tau^{z_1}_\text{diss} &= \frac{1}{\mu \kappa \big( 1 + 4 \frac{\gamma^2}{\kappa^2} \big)}, \quad & \tau^{z_1}_\text{geom} &= 0, \\
\tau^{z_2}_\text{diss} &= \frac{1}{2 \mu \kappa \big( 1 + \frac{5}{2} \frac{\gamma^2}{\kappa^2} \big)}, \quad &\tau^{z_2}_\text{geom} &= \frac{1}{2 \mu \kappa}, \\
\tau^{z_3}_\text{diss} &= \frac{1}{2 \mu \kappa \big( 1 + 2 \frac{\gamma^2}{\kappa^2} \big)}, \quad &\tau^{z_3}_\text{geom} &= \frac{1}{2 \mu \kappa} ,
\end{align} \label{parabolic-speed-limit}%
\end{subequations}
which are shown in Fig.~\ref{fig-parabolic}b)-d).
Here we used $\chi_\perp(\bm{x}) = x_1^2 + x_2^2$ in \eqref{speed-limit-geometric} to obtain a definite lower bound.
Comparing this to the actual values in \eqref{parabolic-correlation}, we see that the geometric speed limit is tight in the strong driving limit for all observables.
For $z_1 = x_1$, the dissipative speed limit captures the qualitative behavior very well, since the correlation time can be reduced arbitrarily much by increasing the driving strength.
For $z_2 = x_1^2$, on the other hand, the decrease in the correlation time only occurs for small to moderate driving, whereas for strong driving, we observe saturation to the geometric speed limit.
For $z_3 = x_1^2 + x_2^2$, the behavior is described by the geometric speed limit for any driving strength, since $\grad z_3(\bm{x}) \cdot \bm{\nu}_\text{st}(\bm{x}) = 0$ and thus \eqref{speed-limit-geometric-observable} applies.

\begin{figure}
\includegraphics[width=.47\textwidth]{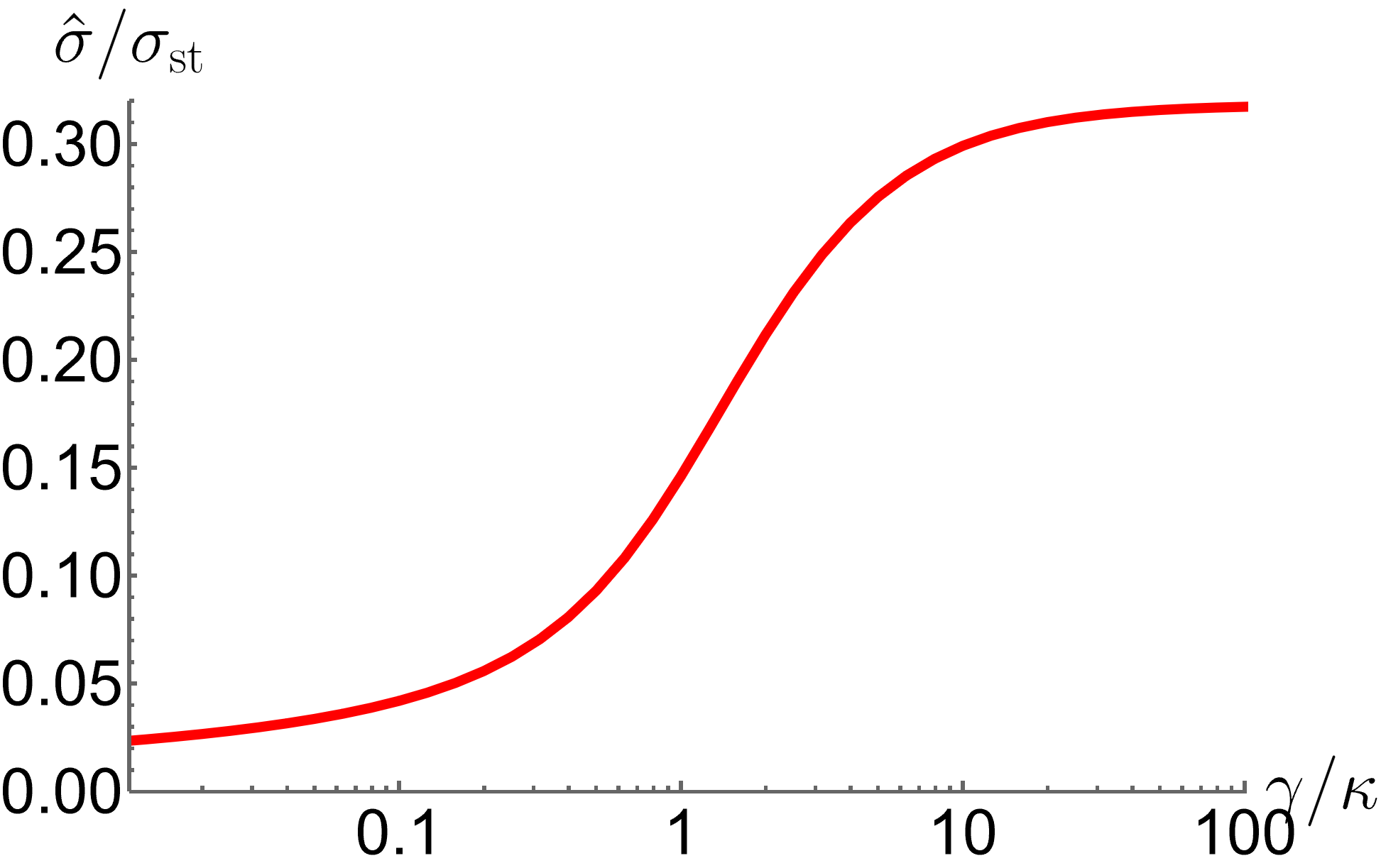}
\caption{The ratio of the lower bound $\hat{\sigma}$ on the entropy production rate given by \eqref{correlation-uncertainty-2} and the actual value, as a function of the driving strength.}
\label{fig-parabolic-entropy}
\end{figure}
Finally, we apply \eqref{correlation-uncertainty-2} to obtain an estimate on the dissipation from a measurement of $x_1$.
Since \eqref{correlation-uncertainty-2} requires a bounded function, we choose $\chi(\bm{x}) = x_1$ for $-\Delta/2 \leq x_1 \leq \Delta/2$, $\chi(\bm{x}) = -\Delta/2$ for $x_1 < -\Delta/2$ and $\chi(\bm{x}) = \Delta/2$ for $x_1 > \Delta/2$, which corresponds to introducing a cutoff value $\pm \Delta/2$ on $x_1$.
For this choice, the right-hand side of \eqref{correlation-uncertainty-2} can be evaluated analytically in terms of $\Delta$, see \eqref{self-averaging-tur-parabolic} in the appendix.
Numerically maximizing this expression with respect to $\Delta$ yields an estimate $\hat{\sigma}$ for the steady-state entropy production, which is shown in Fig.~\ref{fig-parabolic-entropy}.
The estimate is positive for any driving strength, allowing us to conclude that the system is out of equilibrium.
For strong driving, the estimate tends to $\frac{1}{\pi} \approx 0.32$ of the true value and thus reproduces a sizable fraction of the actual dissipation.
We stress that this estimate relies only on a measurement of the fluctuations of $x_1$, while there is no current in $x_1$-direction, prohibiting the application of the thermodynamic uncertainty relation if only $x_1$ can be observed.
Moreover, it can be shown (see section \ref{sec-parabolic-path} of the appendix) that the joint probability density for $x_1$, $p(x_1,\tau;x_1',0)$, is symmetric under exchanging $x_1$ and $x_1'$, so no dissipation can be inferred from the time-reversal properties of the observable.

\textit{Discussion.}
We demonstrated that the self-averaging of observables in the steady state has to obey certain speed limits.
For equilibrium systems, the speed limit takes the form of a tradeoff between short- and long-time fluctuations, which represents the fact that an equilibrium system can only explore its configuration space via diffusive motion.
On the other hand, for nonequilibrium systems, the presence of directed currents generally accelerates the dynamics, allowing for faster self-averaging at the cost of incurring dissipation.

The complimentary speed limits \eqref{speed-limit-sigma} and \eqref{speed-limit-geometric} for nonequilibrium systems highlight different aspects of this phenomenon:
A given speedup requires a minimal amount of dissipation, enabling us to estimate dissipation from a measurement of the violation of the equilibrium speed limit.
Nevertheless, driving the system ever further from equilibrium will not generally result in an arbitrary acceleration of the dynamics.
In an equilibrium system, the slowest timescale, often corresponding to a large-scale physical process, limits how fast the system can explore its configuration space.
Driving the system out of equilibrium in an appropriate manner can speed up this particular process. 
However, at some point another timescale associated with another, smaller scale, process unaffected by the driving will become the limiting factor.
Increasing the magnitude of the driving will then not result in further acceleration, unless the geometric structure of the driving is changed to also affect smaller scales.

We remark that driving a system out of equilibrium to speed up its relaxation has a concrete application in the form of so-called non-reversible sampling in Monte-Carlo simulations \cite{Hwa93,Hwa05,Suw10,Tur11,Ich13,Hwa15,Dun16,Dun17,Cog21}.
There, a perturbation of a given equilibrium system is constructed to preserve the steady state distribution while breaking detailed balance, speeding up both the convergence towards the steady state as well as the sampling of the configuration space in the steady state.
Our speed limits demonstrate that the effectiveness of this approach is constrained both by the strength of the driving as well as its geometric structure.

From the point of view of thermodynamic inference (see Ref.~\cite{Sei19} for a recent perspective), \eqref{correlation-uncertainty} provides a way of estimating dissipation from the measurement of the fluctuations of a time-symmetric observable, even in cases where no currents or time-asymmetry of the transition probability can be used to do so.
We speculate that this approach may be useful in models of active matter \cite{Enc11,Pot12}, where time-asymmetry often manifests only in hidden degrees of freedom and cannot be observed directly from the particles' trajectories, while it may still affect the correlation times of trajectory-dependent observables.

\begin{acknowledgments}
A.~D.~is supported by JSPS KAKENHI (Grant No. 19H05795, and 22K13974). J.~G.~B.’s research was conducted within the Econophysics \& Complex Systems Research Chair, under the aegis of the Fondation du Risque, the Fondation de l’Ecole polytechnique, the Ecole polytechnique and Capital Fund Management. J.~G.~B.~also acknowledges funding from the JSPS summer programme. S.~S.~is supported by JSPS KAKENHI (Grant No. 19H05795, 20K20425, and 22H01144). 
\end{acknowledgments}

\appendix

\onecolumngrid

\section{Derivation of the variational formula for the correlation time} \label{sec-variational}
As in the main text, we first consider the overdamped Langevin dynamics
\begin{align}
\dot{\bm{x}}(t) = \bm{a}(\bm{x}(t)) + \bm{G} \bm{\xi}(t) \label{app-langevin}
\end{align}
with drift vector field $\bm{a}(\bm{x})$ and full-rank matrix $\bm{G}$.
The variational formula for the correlation time follows from a variational formula for the cumulant generating function of time-intergrated observables.
We consider the time-integral of some observable $z(\bm{x})$,
\begin{align}
Z_\tau = \int_0^\tau dt \ z(\bm{x}(t)) \label{observable},
\end{align}
where $\bm{x}(t)$ obeys the Langevin equation \eqref{app-langevin}.
The fluctuations of the time-integrated observable $Z_\tau$ can be characterized using the cumulant generating function
\begin{align}
K^Z(h) = \ln \Av{e^{h Z_\tau}} .
\end{align}
For a diffusive dynamics in the steady state, the cumulants of $Z_\tau$ grow asymptotically linear in time, and we define the scaled cumulant generating function 
\begin{align}
k^Z(h) = \lim_{\tau \rightarrow \infty} \frac{K_Z(h)}{\tau} .
\end{align}
For time-integrated currents, a variational formula for the cumulant generating function has been derived in Ref.~\cite{Nem11} and subsequently generalized in Ref.~\cite{Dec17}.
This derivation generalizes in a straightforward manner to time-integrated observables of the form \eqref{observable}, where we have
\begin{align}
k^Z(h) &= \sup_y \Phi^z[y] \qquad \text{with} \label{cgf-variational} \qquad
\Phi^z[y] = h \av{z}^y_\text{st} - \frac{1}{4} \Av{\bm{y} \cdot \bm{B}^{-1} \bm{y}}_\text{st}^y ,
\end{align}
where $\bm{B} = \bm{G} \bm{G}^\text{T}/2$ is the positive definite diffusion matrix.
Here, $\av{\ldots}_\text{st}^y$ denotes an average with respect to the steady state probability density obtained from a Langevin dynamics with additional drift vector $\bm{y}(\bm{x})$,
\begin{align}
\dot{\bm{x}}(t) = \bm{a}(\bm{x}(t)) + \bm{y}(\bm{x}(t)) + \bm{G} \bm{\xi}(t) \label{langevin-mod} .
\end{align}
In other words, $p_\text{st}^y(\bm{x})$ is the steady-state solution of the Fokker-Planck equation
\begin{align}
0 = -\grad \cdot \Big( \big( \bm{a}(\bm{x}) + \bm{y}(\bm{x}) + \bm{B} \grad \big) p_\text{st}^y(\bm{x}) \Big) \label{fpe-mod} .
\end{align}
Evaluating $\Phi^z[y]$ in \eqref{cgf-variational} thus involves two steps: 
First, choose a drift vector $\bm{y}(\bm{x})$ and compute the (unique) corresponding steady state probability density.
Second, use the probability density to compute the average in \eqref{cgf-variational}.
This procedure has to be repeated for different drift vectors until the maximum is achieved.
However, we can also consider the following, equivalent procedure: 
We first fix the probability density $p_\text{st}^y(\bm{x})$.
Then, \eqref{fpe-mod} determines the allowed drift vectors $\bm{y}(\bm{x})$ that lead to this steady state.
Note that this identification is not unique, since, generally, infinitely many choices of the drift vector yield the same steady state.
However, finding the drift vector that maximizes $\Phi^z[y]$ while yielding the desired steady state corresponds to a convex minimization problem and thus has a unique solution.
Thus, for any given steady state $p_\text{st}^y(\bm{x})$, we can find a unique drift vector $\bm{y}^*(\bm{x})$ that maximizes $\Phi^z[y]$.
Finally, we maximize the resulting expression over all probability densities $p_\text{st}^y(\bm{x})$, re-obtaining \eqref{cgf-variational}.
Specifically, maximizing $\Phi^z[y]$ under the constraint \eqref{fpe-mod} yields
\begin{gather}
\bm{y}^*(\bm{x}) = 2 \bm{B} \grad \phi^*(\bm{x}) \qquad \text{with}  \\
0 = \grad \cdot \Big( p_\text{st}(\bm{x}) e^{\chi(\bm{x})} \big( \bm{\nu}_\text{st}(\bm{x}) + 2 \bm{B} \grad \phi^*(\bm{x}) - \bm{B} \grad \chi(\bm{x}) \big) \Big) \n ,
\end{gather}
where $\phi^*(\bm{x})$ is a Lagrange multiplier.
Here, we introduced the function $\chi(\bm{x}) = \ln (p_\text{st}^y(\bm{x})/p_\text{st}(\bm{x}))$ and used the definition of the steady-state local mean velocity,
\begin{align}
\bm{\nu}_\text{st}(\bm{x}) = \bm{a}(\bm{x}) - \bm{B} \grad \ln p_\text{st}(\bm{x}) ,
\end{align}
which satisfies the steady-state Fokker-Planck equation
\begin{align}
0 = - \grad \cdot \big( \bm{\nu}_\text{st}(\bm{x}) p_\text{st}(\bm{x}) \big) \label{steady-state} .
\end{align}
We further introduce the function $\eta^*(\bm{x}) = 2 \phi^*(\bm{x}) - \chi(\bm{x})$, which results in
\begin{gather}
\Phi^z[y^*] = h \av{z e^{\chi}}_\text{st} - \frac{1}{4}\Av{\grad (\eta^* + \chi) \cdot \bm{B} \grad (\eta^* + \chi) e^{\chi}}_\text{st} \nn
\text{with} \qquad 0 = \grad \cdot \Big( p_\text{st}(\bm{x}) e^{\chi(\bm{x})} \big( \bm{\nu}_\text{st}(\bm{x}) + \bm{B} \grad \eta^*(\bm{x}) \big) \Big) \label{optimal-drift}  .
\end{gather}
Multiplying \eqref{optimal-drift} by $\eta^*(\bm{x})$ and $\chi(\bm{x})$, respectively and integrating, we obtain the conditions
\begin{align}
\Av{\grad \eta^* \cdot \bm{B} \grad \eta^* e^{\chi}}_\text{st} &= - \Av{\grad \eta^* \cdot \bm{\nu}_\text{st} e^{\chi}}_\text{st} \label{eta-chi-conditions} \\
\Av{\grad \chi \cdot \bm{B} \grad \eta^* e^{\chi}}_\text{st} &= - \Av{\grad \chi \cdot \bm{\nu}_\text{st} e^{\chi}}_\text{st} = 0 \n ,
\end{align}
where we integrated by parts and used \eqref{steady-state},
\begin{align}
\av{\grad \chi \cdot \bm{\nu}_\text{st} e^{\chi}}_\text{st} &= \av{\grad e^\chi \cdot \bm{\nu}_\text{st}}_\text{st} = - \int d\bm{x} \ e^{\chi(\bm{x})} \grad \cdot \big( \bm{\nu}_\text{st}(\bm{x}) p_\text{st}(\bm{x}) \big) = 0
\end{align}
in the second equation.
Using this, we can write
\begin{align}
\Phi^z[y^*] &= h \av{z e^{\chi}}_\text{st} \label{functional-eta-chi} - \frac{1}{4} \Big(\Av{\grad \eta^*\cdot \bm{B} \grad \eta^* e^{\chi}}_\text{st} + \Av{\grad \chi \cdot \bm{B} \grad \chi e^{\chi}}_\text{st} \Big) .
\end{align}
The crucial realization needed to proceed further is that the equation determining $\eta^*(\bm{x})$ in \eqref{optimal-drift} is also obtained as the Euler-Lagrange equation of the convex minimization problem
\begin{align}
\inf_\eta \bigg[ &\Av{ \big(\grad \eta \cdot \bm{B} \grad \eta + 2 \grad \eta \cdot \bm{\nu}_\text{st} \big) e^{\chi}}_\text{st} \bigg] \label{eta-variational} 
\end{align}
That is, the minimizer of \eqref{eta-variational} is $\eta^*(\bm{x})$ determined by \eqref{optimal-drift}.
Consequently, we have, using \eqref{eta-chi-conditions},
\begin{align}
\inf_\eta \bigg[ &\Av{ \big(\grad \eta \cdot \bm{B} \grad \eta + 2 \grad \eta \cdot \bm{\nu}_\text{st} \big) e^{\chi}}_\text{st} \bigg] = - \Av{\grad \eta^* \cdot \bm{B} \grad \eta^* e^{\chi}}_\text{st} .
\end{align}
This is precisely the second term in \eqref{functional-eta-chi}, which allows us to write
\begin{align}
\Phi^z[y^*] = \inf_\eta \bigg[ &h \av{z e^{\chi}}_\text{st} - \frac{1}{4} \Av{\grad \chi\cdot \bm{B} \grad \chi e^{\chi}}_\text{st} \label{functional-eta-chi-2} + \frac{1}{4} \Av{ \big(\grad \eta \cdot \bm{B} \grad \eta + 2 \grad \eta \cdot \bm{\nu}_\text{st} \big) e^{\chi}}_\text{st} \bigg]  .
\end{align}
Now all that is left to recover the generating function is to maximize with respect to the probability density $p_\text{st}^y(\bm{x})$.
This is equivalent to maximizing with respect to $\chi(\bm{x})$ under the condition that $e^{\chi}(\bm{x}) p_\text{st}(\bm{x})$ is a normalized probability density, which can be satisfied by subtracting the logarithm of the partition function $\av{e^\chi}_\text{st}$ from $\chi(\bm{x})$.
We obtain
\begin{align}
k^Z(h) = &\sup_\chi \inf_\eta \Bigg[ \frac{1}{\av{e^{\chi}}_\text{st}} \bigg( h \Av{z e^{\chi}}_\text{st} + \Av{\grad \eta \cdot \bm{\nu}_\text{st} e^{\chi}}_\text{st} \label{cgf-variational-final} + \Av{\grad \eta \cdot \bm{B} \grad \eta e^{\chi}}_\text{st} - \frac{1}{4}\Av{\grad \chi \cdot \bm{B} \grad \chi e^{\chi}}_\text{st} \bigg) \Bigg] .
\end{align}
Importantly, this expression does not involve any additional constraints on $\eta(\bm{x})$ and $\chi(\bm{x})$, so that arbitrary choices of either function can be used to obtain upper or lower bounds on the generating function.
The expression for the variance of the time-average is obtained by first noting that, for small $h$,
\begin{align}
k^Z(h) \simeq h \av{z}_\text{st} + \frac{h^2}{2} \tau \text{Var}(\bar{z}) + O(h^3) .
\end{align}
Setting $\chi(\bm{x}) = h \chi_0(\bm{x})$ and $\eta(\bm{x}) = h \eta_0(\bm{x})$, where $\chi_0(\bm{x})$ and $\eta_0(\bm{x})$ are assumed to be of order 1 in the limit $h \rightarrow 0$, we can expand the right-hand side of \eqref{cgf-variational-final} in powers of $h$,
\begin{align}
&\frac{1}{\av{e^{h \chi_0}}_\text{st}} \bigg( h \Av{z e^{h \chi_0}}_\text{st} + h \Av{\grad \eta_0 \cdot \bm{\nu}_\text{st} e^{h \chi_0}}_\text{st} + h^2 \Av{\grad \eta_0 \cdot \bm{B} \grad \eta_0 e^{h \chi}}_\text{st} - \frac{h^2}{4}\Av{\grad \chi_0 \cdot \bm{B} \grad \chi_0 e^{h \chi_0}}_\text{st} \bigg) \\
&\hspace{1cm} \simeq h \av{z}_\text{st} + h^2 \bigg( \text{Cov}_\text{st}(z,\chi_0) + \av{\chi_0 \grad \eta_0 \cdot \bm{\nu}_\text{st}}_\text{st} + \Av{\grad \eta_0 \cdot \bm{B} \grad \eta_0}_\text{st}  - \frac{1}{4}\Av{\grad \chi_0 \cdot \bm{B} \grad \chi_0}_\text{st} \bigg) + O(h^3) . \n 
\end{align}
The term linear in $h$ cancels and we obtain by comparing the coefficients of the quadratic terms,
\begin{align}
\frac{\tau}{2} \text{Var}(\bar{z}) &\simeq \sup_\chi \inf_\eta \Bigg[  \text{Cov}_\text{st}(z,\chi) + \av{\chi \grad \eta \cdot \bm{\nu}_\text{st}}_\text{st} + \Av{\grad \eta \cdot \bm{B} \grad \eta}_\text{st} - \frac{1}{4}\Av{\grad \chi \cdot \bm{B} \grad \chi}_\text{st} \Bigg],
\end{align}
where we removed the subscript 0.
The final step is to rescale $\chi(\bm{x}) \rightarrow \alpha \chi(\bm{x})$ and $\eta(\bm{x}) \rightarrow \beta \eta(\bm{x})$ and solve the quadratic optimization problem for the parameters $\alpha$ and $\beta$, which yields
\begin{align}
\frac{\tau}{2} \text{Var}(\bar{z}) &\simeq \sup_\chi \inf_\eta \Bigg[  \frac{\text{Cov}_\text{st}(z,\chi)^2}{\av{\grad \chi \cdot \bm{B} \grad \chi}_\text{st} + \frac{\av{\chi \grad \eta \cdot \bm{\nu}_\text{st}}_\text{st}^2}{\av{\grad \eta \cdot \bm{B} \grad \eta}_\text{st}}} \Bigg] .
\end{align}
Recalling the definition of the correlation time,
\begin{align}
\frac{\text{Var}(\bar{z}_\tau)}{2 \text{Var}_\text{st}(z)} \simeq \frac{\tau^z}{\tau} \label{correlation-time} ,
\end{align}
we obtain \eqref{correlation-variational} of the main text,
\begin{align}
\tau^z = \max_{\chi} \Bigg[ \frac{\frac{\text{Cov}_\text{st}(z,\chi)^2}{\text{Var}_\text{st}(z)}}{\av{\grad \chi \cdot \bm{B} \grad \chi}_\text{st} + \max_{\eta} \Big[ \frac{\av{\chi \grad \eta \cdot \bm{\nu}_\text{st}}_\text{st}^2}{\av{\grad \eta \cdot \bm{B} \grad \eta}_\text{st}} \Big]} \Bigg] \label{app-correlation-variational} .
\end{align}

\section{Short-time fluctuations of the displacement} \label{sec-displacement}
For a state-dependent observable $z(\bm{x}(t))$, we consider its displacement $dz(t) = z(\bm{x}(t+dt)) - z(\bm{x}(t))$, that is, how much its value changes due to a change in the configuration $\bm{x}(t)$ of the system.
According to the chain rule of stochastic calculus, the displacement can be expressed in terms of the Stratonovich product \cite{Gar02},
\begin{align}
dz(t) = \grad z(\bm{x}(t)) \circ d\bm{x}(t) ,
\end{align}
where the displacement $d\bm{x}(t)$ of the configuration is given by the Langevin \eqref{langevin},
\begin{align}
d\bm{x}(t) = \bm{a}(\bm{x}(t)) dt + \bm{G} d\bm{w}(t),
\end{align}
where $d\bm{w}(t)$ is a vector of increments of $d$ mutually independent Wiener processes.
We can transform the Stratonovich product into an Ito product by using the Ito formula
\begin{align}
dz(t) = \grad z(\bm{x}(t)) \cdot \big( \bm{a}(\bm{x}(t)) dt + \bm{G} d\bm{w}(t) \big) + \text{tr}\big(\bm{B} \bm{\mathcal{H}}^z(\bm{x}(t)) \big) dt ,
\end{align}
where tr denotes the trace and $\bm{\mathcal{H}}^z(\bm{x})$ is the Hessian matrix of $z(\bm{x})$.
We then have
\begin{align}
\Av{dz(t)^2} \simeq 2 \Av{\grad z \cdot \bm{B} \grad z}_\text{st} dt + O(dt dw) + O(dt^2) ,
\end{align}
where we used the covariance of the Wiener increments,
\begin{align}
\av{dw_i(t) dw_j(t)} = \delta_{ij} dt .
\end{align}
Since $\av{dz(t)} \simeq O(dt)$, the above result is to leading order in $dt$ equal to the variance and we have
\begin{align}
\lim_{dt \rightarrow 0} \bigg( \frac{\text{Var}(dz(t))}{2 dt} \bigg) = \Av{\grad z \cdot \bm{B} \grad z}_\text{st} .
\end{align}

\section{Equilibrium tradeoff for underdamped Langevin dynamics} \label{sec-underdamped}
We consider the underdamped Langevin dynamics with linear Stokes friction
\begin{align}
\dot{\bm{x}}(t) = \bm{v}(t) \qquad \text{and} \qquad \bm{m} \dot{\bm{v}}(t) = \bm{f}(\bm{x}(t)) - \bm{\gamma} \bm{v}(t) + \sqrt{2 \bm{\gamma} T} \bm{\xi}(t) \label{langevin-under} .
\end{align}
Here, $\bm{f}(\bm{x})$ is the force acting on the system and $\bm{m}$ and $\bm{\gamma}$ are diagonal matrices containing the masses and friction coefficients associated with the individual particles.
The steady state of this system determined by the Kramers-Fokker-Planck equation
\begin{align}
0 = \Big( - \bm{v} \cdot \grad_x - \bm{m}^{-1} \grad_v \cdot \big( \bm{f}(\bm{x}) - \bm{\gamma} \bm{v} - T \bm{m}^{-1} \bm{\gamma} \grad_v \big) \Big) p_\text{st}(\bm{x},\bm{v}) \label{kramers} .
\end{align}
We can write this as
\begin{align}
0 = -\grad \cdot \Big( \bm{\omega}_\text{st}(\bm{x},\bm{v}) p_\text{st}(\bm{x},\bm{v}) \Big) \qquad \text{with} \qquad \bm{\omega}_\text{st}(\bm{x},\bm{v}) = \begin{pmatrix}
\bm{v} \\
\bm{m}^{-1} \big( \bm{f}(\bm{x}) - \bm{\gamma} \bm{v} - T \bm{m}^{-1} \bm{\gamma} \grad_v \ln p_\text{st}(\bm{x},\bm{v}) 
\end{pmatrix} , \label{kramers-meanvel}
\end{align}
where we defined $\grad = (\grad_x,\grad_v)$ and $\bm{\omega}_\text{st}(\bm{x},\bm{v})$ plays the role of the local mean velocity in the overdamped case.
Compared to \eqref{fpe-mod}, the diffusion matrix of this system is singular and thus we cannot write down the analog of \eqref{cgf-variational} immediately.
However, we can introduce the auxiliary dynamics
\begin{align}
\dot{\bm{x}}(t) = \bm{v}(t) + \epsilon \grad_x \ln p_\text{st}(\bm{x}(t),\bm{v}(t)) + \sqrt{2 \epsilon} \bm{\zeta}(t)  \qquad \text{and} \qquad \bm{m} \dot{\bm{v}}(t) = \bm{f}(\bm{x}(t)) - \bm{\gamma} \bm{v}(t) + \sqrt{2 \bm{\gamma} T} \bm{\xi}(t) ,
\end{align}
where $\epsilon$ is a parameter, $\bm{\xi}(t)$ and $\bm{\zeta}(t)$ are mutually independent white noises and $p_\text{st}(\bm{x},\bm{v})$ is the solution of \eqref{kramers}.
Formally, this can be viewed as an overdamped dynamics with steady-state condition
\begin{align}
0 = \Big( - \bm{v} \cdot \grad_x - \grad_x \cdot \big( \epsilon [ \grad_x \ln p_\text{st}(\bm{x},\bm{v})] - \epsilon \grad_x \big) - \bm{m}^{-1} \grad_v \big( \bm{f}(\bm{x}) - \bm{\gamma} \bm{v} - T \bm{m}^{-1} \bm{\gamma} \grad_v \big) \Big) p_\text{st}(\bm{x},\bm{v}) .
\end{align}
It is obvious that $p_\text{st}(\bm{x},\bm{v})$ is the solution with the same value for $\bm{\omega}_\text{st}(\bm{x},\bm{v})$ as in \eqref{kramers-meanvel}.
However, in contrast to \eqref{kramers}, the diffusion matrix is now positive definite and we can apply \eqref{cgf-variational} to the auxiliary dynamics.
For the correlation time of some observable $z(\bm{x},\bm{v})$, we then obtain, in analogy to \eqref{app-correlation-variational}
\begin{align}
\tau^z = \sup_{\chi} \Bigg[ \frac{\frac{\text{Cov}_\text{st}(z,\chi)^2}{\text{Var}_\text{st}(z)}}{\Av{ \grad_v \chi \cdot T \bm{\gamma} \bm{m}^{-2} \grad_v \chi}_\text{st} + \epsilon \Av{\Vert\grad_x \chi \Vert^2}_\text{st} + \sup_\eta \Big[ \frac{\Av{\chi \grad \eta \cdot \bm{\omega}}_\text{st}^2}{\Av{ \grad_v \eta \cdot T \bm{\gamma} \bm{m}^{-2} \grad_v \eta}_\text{st} + \epsilon \Av{\Vert \grad_x \eta \Vert^2}_\text{st}} \Big]  } \Bigg] \label{correlation-variational-under} .
\end{align}
Provided that the limit $\epsilon \rightarrow 0$ is sufficiently regular, we can then expect to obtain the correlation time corresponding to \eqref{langevin-under} in this limit.
Note that the $\epsilon$-regularization in \eqref{correlation-variational-under} is generally required, since otherwise, we could choose functions $\chi(\bm{x},\bm{v})$ and $\eta(\bm{x},\bm{v})$ with a very irregular dependence on $\bm{x}$, which lead to trivial results for the maximizer.
Physically, this represents the fact that not every arbitrary probability density $q(\bm{x},\bm{v})$ can be obtained as a steady-state solution of \eqref{kramers} for some choice of the force $\bm{f}(\bm{x})$, in contrast to the overdamped case, where a drift vector leading to a specific probability density can always be found.
For position-dependent observables in equilibrium, however, this does not cause any issues.
Specifically, if $\bm{f}(\bm{x}) = - \grad U(\bm{x})$, then the solution to \eqref{kramers} is given by
\begin{align}
p_\text{eq}(\bm{x},\bm{v}) = \frac{e^{-\frac{U(\bm{x})}{T}}}{\int d\bm{x} \ e^{-\frac{U(\bm{x})}{T}}} \sqrt{\frac{\det(\bm{m})}{(2 \pi T)^d}} e^{-\frac{1}{2 T} \bm{v} \cdot \bm{m} \bm{v}} \label{bg-under} .
\end{align}
We focus on observables that only depend on the position, $z(\bm{x},\bm{v}) = z(\bm{x})$.
Choosing $\chi(\bm{x},\bm{v}) = z(\bm{x})$ and setting $\epsilon = 0$, we have the lower bound
\begin{align}
\tau_\text{eq}^z \geq \text{Var}_\text{eq}(z) \inf_\eta \Bigg[ \frac{\Av{ \grad_v \eta \cdot T \bm{\gamma} \bm{m}^{-2} \grad_v \eta}_\text{eq}}{\Av{\eta \grad_x z \cdot \bm{v}}_\text{eq}^2} \Bigg],
\end{align}
where we integrated by parts in the denominator using \eqref{kramers-meanvel}.
The Euler-Lagrange equation corresponding to the minimization over $\eta(\bm{x},\bm{v})$ reads
\begin{align}
\grad_v \cdot \big( p_\text{eq}(\bm{x},\bm{v}) T \bm{\gamma} \bm{m}^{-2} \grad_v \eta(\bm{x},\bm{v}) \big) + p_\text{eq}(\bm{x},\bm{v}) \grad_x z(\bm{x}) \cdot \bm{v} = 0.
\end{align}
Using \eqref{bg-under}, it can be checked that the solution is given by
\begin{align}
\eta(\bm{x},\bm{v}) = \bm{m} \bm{\gamma}^{-1} \grad_x z(\bm{x}) \cdot \bm{v} .
\end{align}
Plugging this in, we obtain the lower bound
\begin{align}
\tau_\text{eq}^z \geq \frac{\text{Var}_\text{eq}(z)}{\Av{\grad_x z \cdot T \bm{\gamma}^{-1} \grad_x z}_\text{eq}} .
\end{align}
In fact, repeating the above calculation with an arbitrary position-dependent function $\chi(\bm{x},\bm{v}) = \chi(\bm{x})$ yields the bound,
\begin{align}
\tau_\text{eq}^z \geq \sup_{\chi} \Bigg[ \frac{\frac{\text{Cov}_\text{eq}(z,\chi)^2}{\text{Var}_\text{eq}(z)}}{\Av{\grad_x \chi \cdot T \bm{\gamma}^{-1} \grad_x \chi}_\text{eq}} \Bigg] .
\end{align}
Comparing this to the overdamped result, \eqref{speed-limit-eq} in the main text,
\begin{align}
\tau_\text{eq,od}^z = \sup_{\chi} \Bigg[ \frac{\frac{\text{Cov}_\text{eq}(z,\chi)^2}{\text{Var}_\text{eq}(z)}}{\Av{\grad_x \chi \cdot \bm{B} \grad_x \chi}_\text{eq}} \Bigg] ,
\end{align}
we therefore conclude that the equilibrium correlation time in the underdamped system is always larger than the correlation time in the overdamped system with the same steady state position-density and diffusion matrix $\bm{B} = T \bm{\gamma}^{-1}$.
Interestingly, even though in overdamped systems, oscillatory motion induced by the irreversible flows decreases the correlation time, the oscillatory motion due to finite mass generally has the opposite effect.
In order to interpret the above as a tradeoff between short- and long-time fluctuations, we note that the displacement $dz = z(\bm{x}(t+dt)) - z(\bm{x}(t))$ exhibits ballistic scaling in the underdamped case,
\begin{align}
\text{Var}(dz) \simeq T \Av{\grad_x z \cdot \bm{m}^{-1} \grad_x z}_\text{eq} dt^2 = \text{Var}(dz) \simeq T \Av{\grad_x z \cdot \bm{\gamma}^{-1} \bm{\gamma} \bm{m}^{-1} \grad_x z}_\text{eq} dt^2 .
\end{align}
Defining the (maximal) thermalization time
\begin{align}
\tau_\text{th} = \max_{i} \bigg( \frac{m_i}{\gamma_i} \bigg) ,
\end{align}
we have the bound
\begin{align}
\lim_{dt \rightarrow 0}  \bigg( \frac{\text{Var}(dz)}{dt^2} \bigg) \geq \frac{1}{t_\text{th}} T \Av{\grad_x z \cdot \bm{\gamma}^{-1} \grad_x z}_\text{eq} .
\end{align}
Therefore, we can write
\begin{align}
\lim_{\tau \rightarrow \infty} \bigg( \frac{\tau}{2} \text{Var}(\bar{z}_\tau) \bigg) \lim_{dt \rightarrow 0} \bigg( \frac{\text{Var}(dz)}{dt^2} \bigg) \geq \frac{1}{\tau_\text{th}} \text{Var}_\text{eq}(z)^2,
\end{align}
which is the analog of the overdamped tradeoff relation
\begin{align}
\lim_{\tau \rightarrow \infty} \bigg( \frac{\tau}{2} \text{Var}(\bar{z}_\tau) \bigg) \lim_{dt \rightarrow 0} \bigg( \frac{\text{Var}(dz)}{2 dt} \bigg) \geq \text{Var}_\text{eq}(z)^2 .
\end{align}
In both overdamped and underdamped systems, the product of the long-time fluctuations of the time-average and the short-time fluctuations of the displacement thus obey a similar tradeoff relation.
In both cases, the intuition behind the tradeoff is that, since, at long times, the system can only explore its configuration space via reversible diffusion, faster self-averaging requires increasing the magnitude of the short-time fluctuations.

\section{Driven diffusion in a parabolic potential} \label{sec-parabolic}

\subsection{Transition probability and correlation time} \label{sec-parabolic-transition}
Here, we provide the details of the calculation for the illustrative example discussed in the main text.
For a parabolic potential $U(x_1,x_2) = \kappa (x_1^2 + x_2^2)/2$ and driving force $\bm{F}_\text{nc}(x_1,x_2) = \gamma(x_2,-x_1)$, the Langevin equation is
\begin{align}
\dot{x}_1(t) = - \mu \kappa x_1(t) + \mu \gamma x_2(t) + \sqrt{2 \mu T} \xi_1(t) \qquad \text{and} \qquad \dot{x}_2(t) = - \mu \kappa x_2(t) - \mu \gamma x_2(t) + \sqrt{2 \mu T} \xi_2(t) .
\end{align}
The corresponding Fokker-Planck equation is
\begin{align}
\partial_t p_t(x_1,x_2) = \mu \bigg( \partial_{x_1} \Big( \big(\kappa x_1 - \gamma x_2 + T \partial_{x_1} \big) p_t(x_1,x_2) \Big) + \partial_{x_2} \Big( \big(\kappa x_2 + \gamma x_1 + T \partial_{x_2} \big) p_t(x_1,x_2) \Big) \bigg) .
\end{align}
Since the Langevin equation is linear, the corresponding probability density and transition probability density are Gaussian, provided that their initial data is Gaussian.
In particular, they are completely characterized by their average and covariance matrix, which evolve according to the ordinary differential equations
\begin{subequations}
\begin{gather}
d_t \av{x_1}_t = - \mu \kappa \av{x_1}_t + \mu \kappa \av{x_2}_t, \\
d_t \av{x_2}_t = - \mu \kappa \av{x_2}_t - \mu \kappa \av{x_1}_t, \\
d_t \text{Var}_t(x_1) = - 2 \mu \kappa \text{Var}_t(x_1) + 2 \mu \gamma \text{Cov}_t(x_1,x_2) + 2 \mu T, \\
d_t \text{Var}_t(x_2) = - 2 \mu \kappa \text{Var}_t(x_2) - 2 \mu \gamma \text{Cov}_t(x_1,x_2) + 2 \mu T, \\
d_t \text{Cov}_t(x_1,x_2) = - 2\mu \kappa \text{Cov}_t(x_1,x_2) + \mu \gamma \big( \text{Var}_t(x_2) - \text{Var}_t(x_1) \big)  .
\end{gather}
\end{subequations}
The steady state solution is
\begin{align}
p_\text{st}(x_1,x_2) = \frac{\kappa}{2 \pi T} \exp \bigg[-\frac{\kappa}{2 T} \big(x_1^2 + x_2^2 \big) \bigg] \qquad \text{with} \qquad \bm{\nu}_\text{st}(x_1,x_2) = \mu \gamma \begin{pmatrix} x_2 \\ -x_1 \end{pmatrix} \label{steady-state-parabolic} .
\end{align}
We note that the steady state probability density is equal to the equilibrium Boltzmann-Gibbs density in the parabolic potential, independent of the driving strength $\gamma$.
On the other hand, the transition probability density corresponds to choosing $\av{x_1}_0 = y_1$, $\av{x_2}_0 = y_2$ and $\text{Var}_0(x_1) = \text{Var}_0(x_2) = \text{Cov}_0(x_1,x_2)$ as initial conditions, which yields
\begin{subequations}
\begin{gather}
\av{x_1}_t = e^{-\mu \kappa t} \big( y_1 \cos(\mu \gamma t) - y_2 \sin(\mu \gamma t) \big), \\
\av{x_2}_t = e^{-\mu \kappa t} \big( y_2 \cos(\mu \gamma t) + y_1 \sin(\mu \gamma t) \big), \\
\text{Var}_t(x_1) = \frac{T}{\kappa} \big( 1 - e^{-2 \mu \kappa t} \big) = \text{Var}_t(x_2), \\
\text{Cov}_t(x_1,x_2) = 0 ,
\end{gather}
\end{subequations}
and thus the transition probability density
\begin{align}
p_t(x_1,x_2 \vert y_1,y_2) = \frac{\kappa}{2 \pi T ( 1 - e^{-2 \mu \kappa t}) } \exp \Bigg[ - \frac{\kappa}{2 T(1 - e^{-2 \mu \kappa t})} \bigg( \Big( &x_1 - e^{-\mu \kappa t} \big( y_1 \cos(\mu \gamma t) - y_2 \sin(\mu \gamma t) \big) \Big)^2 \label{transition-parabolic} \\
& + \Big( x_2 - e^{-\mu \kappa t} \big( y_2 \cos(\mu \gamma t) + y_1 \sin(\mu \gamma t) \big) \Big)^2 \bigg) \Bigg] . \n
\end{align}
Using these results, the Green-Kubo formula for the correlation time, \eqref{green-kubo} of the main text,
\begin{align}
\tau^z = \int_0^\infty dt \ \frac{\text{Cov}(z(\bm{x}(t)),z(\bm{x}(0)))}{\text{Var}_\text{st}(z)}, \label{green-kubo-tau}
\end{align}
can be evaluated explicitly for any observable $z(x_1,x_2)$ that is polynomial in $x_1$ and $x_2$ by computing Gaussian integrals.
For example, for $z^1 = x_1$, we have
\begin{align}
\text{Cov}(x_1(t),x_1(0)) &= \int dx_1 \int dx_2 \int dy_1 \int dy_2 \ x_1 y_1 \big( p_t(x_1,x_2 \vert y_1,y_2) - p_\text{st}(x_1,x_2) \big) p_\text{st}(y_1,y_2) = \frac{T}{\kappa} e^{-\mu \kappa t} \cos(\mu \gamma t) .
\end{align}
For the correlation time, this yields
\begin{align}
\tau^{z_1} = \frac{1}{\mu \kappa  \big(1 + \frac{\gamma^2}{\kappa^2} \big)} \label{tau-z1}.
\end{align}
The calculation for the observables $z_2 = x_1^2$ and $z_3 = x_1^2 + x_2^2$ discussed in the main text proceeds in the same way.
For later use, we particularly note that
\begin{align}
\tau^{z_3} = \frac{1}{2 \mu \kappa} \label{tau-z3}.
\end{align}

\subsection{Intrinsic correlation time and eigenvalues} \label{sec-parabolic-intrinsic}
In principle, the intrinsic correlation time can be computed by maximizing \eqref{green-kubo-tau} with respect to the observable $z(\bm{x})$.
There is, however, a simpler way which, as a useful by-product, reveals an interesting connection between the correlation time and the eigenvalues of the Fokker-Planck equation.
First, we note that the maximum over $\eta(\bm{x})$ in \eqref{app-correlation-variational} has an equivalent expression
\begin{align}
\max_{\eta} \bigg[ \frac{\av{\chi \grad \eta \cdot \bm{\nu}_\text{st}}_\text{st}^2}{\av{\grad \eta \cdot \bm{B} \grad \eta}_\text{st}} \bigg] = \max_{\phi} \Big[ 2 \av{\chi \grad \phi \cdot \bm{\nu}_\text{st}}_\text{st} - \Av{\grad \phi \cdot \bm{B} \grad \phi}_\text{st} \Big] \label{maximization-eta} .
\end{align}
Writing $\phi(\bm{x}) = \alpha \eta(\bm{x})$ and then maximizing with respect to the real parameter $\alpha$ on the right-hand side yields the left-hand side.
For the second expression, we can immediately write down the Euler-Lagrange equation
\begin{align}
\grad \cdot \big( \bm{B} p_\text{st}(\bm{x}) \grad \phi^*(\bm{x}) \big) - \grad \chi(\bm{x}) \cdot \bm{\nu}_\text{st}(\bm{x}) p_\text{st}(\bm{x}) = 0 \label{maximizer-eta},
\end{align}
where $\phi^*(\bm{x})$ denotes the maximizer and we used \eqref{steady-state}.
We now introduce the eigenfunctions and eigenvalues of the equilibrium Fokker-Planck operator
\begin{align}
- \grad \cdot \Big( \big( \bm{a}_\text{eq}(\bm{x}) - \bm{B} \grad \big) \psi_k(\bm{x}) \Big) = - \lambda_k \psi_k(\bm{x}) .
\end{align}
Here, $\bm{a}_\text{eq}(\bm{x})$ is the drift vector that yields the same steady state $p_\text{st}(\bm{x})$ as the original dynamics, however, with vanishing local mean velocity and thus vanishing entropy production.
It is given by
\begin{align}
a_\text{eq}(\bm{x}) = \bm{B} \grad \ln p_\text{st}(\bm{x}) .
\end{align}
Since this is an equilibrium dynamics, the Fokker-Planck operator is self-adjoint and its spectrum is real with $\lambda_0 = 0$, $\psi_0(\bm{x}) = p_\text{st}(\bm{x})$ and $\lambda_k > 0$ for $k \geq 1$ \cite{Ris86}.
Writing $\psi_k(\bm{x}) = \chi_k(\bm{x}) p_\text{st}(\bm{x})$, the eigenvalue problem can be written as
\begin{align}
\grad \cdot \big( \bm{B} p_\text{st}(\bm{x}) \grad \chi_k(\bm{x}) \big) = -\lambda_k \chi_k(\bm{x}) p_\text{st}(\bm{x}) \label{eigenvalue-eq} .
\end{align}
We normalize the eigenfunctions to $\av{\chi_k^2}_\text{st} = 1$, so that they satisfy the orthogonality relations
\begin{align}
\av{\chi_k \chi_l}_\text{st} = \delta_{k l} \qquad \text{and} \qquad \av{\grad \chi_k \cdot \bm{B} \grad \chi_l}_\text{st} = \lambda_k \delta_{kl} \label{orthogonality-eq} ,
\end{align}
which can be obtained from \eqref{eigenvalue-eq} after multiplying with $\chi_l(\bm{x})$, integrating and using the fact that $\bm{B}$ is a symmetric matrix.
We write the maximizer in \eqref{maximizer-eta} as a linear combination of the eigenfunctions $\chi_k(\bm{x}$,
\begin{align}
\phi^*(\bm{x}) = \sum_{k=1}^\infty a_k \chi_k(\bm{x}) \label{phi-expansion}.
\end{align}
Note that we have $\chi_0(\bm{x}) \equiv 1$ and so the $k = 0$ term does not contribute to \eqref{maximizer-eta} and we omit it.
Plugging this into \eqref{maximizer-eta}, multiplying by $\chi_l(\bm{x})$ and integrating yields
\begin{align}
- \sum_{k = 1}^\infty a_k \Av{\grad \chi_l \cdot \bm{B} \grad \chi_k}_\text{st} - \Av{\chi_l \grad \chi \cdot \bm{\nu}_\text{st}}_\text{st} = 0.
\end{align}
Using \eqref{orthogonality-eq}, we obtain
\begin{align}
a_l = - \frac{1}{\lambda_l} \Av{\chi_l \grad \chi \cdot \bm{\nu}_\text{st}}_\text{st} .
\end{align}
This determines the coefficients of $\phi^*(\bm{x})$ in the eigenfunction expansion.
On the other hand, in terms of the maximizer \eqref{maximizer-eta}, \eqref{maximization-eta} reads
\begin{align}
\max_{\phi} \Big[ 2 \av{\chi \grad \phi \cdot \bm{\nu}_\text{st}}_\text{st} - \Av{\grad \phi \cdot \bm{B} \grad \phi}_\text{st} \Big] = \Av{\grad \phi^* \cdot \bm{B} \grad \phi^*}_\text{st},
\end{align}
which follows from multiplying \eqref{maximizer-eta} by $\phi^*(\bm{x})$ and integrating.
Plugging in \eqref{phi-expansion}, we find
\begin{align}
\Av{\grad \phi^* \cdot \bm{B} \grad \phi^*}_\text{st} = \sum_{k,l = 1}^\infty a_k a_l \Av{\grad \chi_k \cdot \bm{B} \grad \chi_l}_\text{st} = \sum_{k = 1}^\infty a_k^2 \lambda_k = \sum_{k = 1}^\infty \frac{1}{\lambda_k} \Av{\chi_k \grad \chi \cdot \bm{\nu}_\text{st}}_\text{st}^2 .
\end{align}
We can then write the variational formula for the correlation time as
\begin{align}
\tau^z = \max_{\chi} \Bigg[ \frac{\frac{\text{Cov}_\text{st}(z,\chi)^2}{\text{Var}_\text{st}(z)}}{\av{\grad \chi \cdot \bm{B} \grad \chi}_\text{st} + \sum_{k = 1}^\infty \frac{1}{\lambda_k} \av{\chi_k \grad \chi \cdot \bm{\nu}_\text{st}}_\text{st}^2} \Bigg] .
\end{align}
For the intrinsic correlation time, we obtain in the same way
\begin{align}
\tau^* = \max_{\chi} \Bigg[ \frac{\text{Var}_\text{st}(\chi)}{\av{\grad \chi \cdot \bm{B} \grad \chi}_\text{st} + \sum_{k = 1}^\infty \frac{1}{\lambda_k} \av{\chi_k \grad \chi \cdot \bm{\nu}_\text{st}}_\text{st}^2} \Bigg] .
\end{align}
If we further expand $\chi(\bm{x})$ in terms of the eigenfunctions $\chi_k(\bm{x})$, we obtain the expression
\begin{align}
\tau^* = \max_{b} \Bigg[ \frac{\sum_{k=1}^\infty b_k^2}{\sum_{k = 1}^\infty \lambda_k b_k^2 + \sum_{k,l,m = 1}^\infty \frac{b_l b_m}{\lambda_k} \Av{\chi_k \grad \chi_l \cdot \bm{\nu}_\text{st}}_\text{st} \Av{\chi_k \grad \chi_m \cdot \bm{\nu}_\text{st}}_\text{st}} \Bigg] \label{intrinsic-eigenvalue} .
\end{align}
By expanding in terms of the eigenfunctions of the equilibrium Fokker-Planck operator, we have transformed the maximization over two functions into a maximization over real parameters and evaluating the elements of the skew-symmetric matrix
\begin{align}
\mathcal{M}_{kl} = \av{\chi_k \grad \chi_l \cdot \bm{\nu}_\text{st}}_\text{st} = -\mathcal{M}_{lk} \label{m-matrix}.
\end{align}
Note that, since the maximization over the parameters $b_k$ has the form of a Rayleigh quotient, it corresponds to finding the smallest eigenvalue of the positive definite matrix
\begin{align}
\mathcal{Q}_{lm} = \lambda_l \delta_{lm} + \sum_{k=1}^\infty \frac{1}{\lambda_k} \mathcal{M}_{lk} \mathcal{M}_{mk} .
\end{align}
We remark that the sum over $k$ in the second term of the denominator of \eqref{intrinsic-eigenvalue} only contains positive terms; thus, we get an upper bound on the intrinsic correlation time by truncating it at a finite value of $k_\text{max}$.
This corresponds to replacing the matrix $\bm{\mathcal{Q}}$ by the truncated one
\begin{align}
\tilde{\mathcal{Q}}_{lm} = \lambda_l \delta_{lm} + \sum_{k=1}^{k_\text{max}} \frac{1}{\lambda_k} \mathcal{M}_{kl} \mathcal{M}_{km} .
\end{align}
In particular, neglecting the sum entirely yields the equilibrium correlation time as an upper bound,
\begin{align}
\tau^* \leq \max_{b} \Bigg[ \frac{\sum_{k=1}^\infty b_k^2}{\sum_{k = 1}^\infty \lambda_k b_k^2} \Bigg] = \frac{1}{\lambda_1} = \tau^*_\text{eq} .
\end{align}
For the parabolic potential, the equilibrium Fokker-Planck operator is obtained by simply setting $\gamma = 0$, which corresponds to a trapped particle without driving.
In this case, the eigenvalues and eigenfunctions are straightforward to compute, they are given by
\begin{align}
\lambda_k = k \mu \kappa, \qquad \text{and} \qquad
\chi_{(k,\alpha)}(\bm{x}) = \frac{1}{\sqrt{\alpha!(k-\alpha)!}} \text{He}_{k-\alpha}\bigg( \sqrt{\frac{\kappa}{T}} x_1 \bigg) \text{He}_{\alpha}\bigg( \sqrt{\frac{\kappa}{T}} x_2 \bigg), \n
\end{align}
where $k = 0,1,\ldots$ enumerates the eigenvalues, $\alpha = 0,1,\ldots,k$ accounts for the $k+1$-fold degeneracy of $\lambda_k$ and $\text{He}_n(z)$ denotes the $n$-th Hermite polynomial,
\begin{align}
\text{He}_n(z) = (-1)^n e^{\frac{z^2}{2}} \frac{d^n}{dz^n} e^{-\frac{z^2}{2}} .
\end{align}
Using this, we can compute all entries of the matrix $\bm{\mathcal{M}}$ in \eqref{m-matrix}.
In particular, we find that $\mathcal{M}_{(k,\alpha),(l,\beta)}$ is non-zero only when $k = l$ and $\alpha = \beta \pm 1$,
\begin{align}
\mathcal{M}_{(k,\alpha),(l,\beta)} = \mu \gamma \delta_{kl} \Big( \sqrt{\alpha (k + 1 -\alpha)} \delta_{\alpha,\beta+1} - \sqrt{(\alpha+1) (k - \alpha)} \delta_{\alpha,\beta-1} \Big) .
\end{align}
This means that, if only consider terms with $k = 1$ in \eqref{intrinsic-eigenvalue}, we get the upper bound
\begin{align}
\tau^* &\leq \max_{b} \Bigg[ \frac{\sum_{k=1}^\infty \sum_{\alpha = 0}^k b_{(k,\alpha)}^2}{\mu \kappa \sum_{k = 1}^\infty \sum_{\alpha = 0}^k k b_{(k,\alpha)}^2 + \frac{1}{\mu \kappa} \sum_{\alpha,\beta,\delta = 0}^1  b_{(1,\beta)} b_{(1,\delta)} \mathcal{M}_{(1,\alpha),(1,\beta)} \mathcal{M}_{(1,\alpha),(1,\delta)}} \Bigg] \\
&= \max_{b} \Bigg[ \frac{\sum_{k=1}^\infty \sum_{\alpha = 0}^k b_{(k,\alpha)}^2}{\mu \kappa \sum_{k = 1}^\infty \sum_{\alpha = 0}^k k b_{(k,\alpha)}^2 + \frac{\mu \gamma^2}{\kappa} \big(b_{(1,0)}^2 + b_{(1,1)}^2 \big)} \Bigg] . \n
\end{align}
The maximizer of the above expression depends on the relative size of $\kappa$ and $\gamma$.
For $\gamma < \kappa$, the maximum is attained for $b_{(1,0)}$ and $b_{(1,1)}$ arbitrary but non-zero and $b_{(k,\alpha)} = 0$ for $k \geq 2$.
On the other hand, for $\gamma > \kappa$, the maximum is attained for $b_{(1,0)} = b_{(1,1)} = 0$, while $b_{(2,0)}$, $b_{(2,1)}$ and $b_{(2,2)}$ are arbitrary but non-zero and $b_{(k,\alpha)} = 0$ for $k \geq 3$.
The resulting upper bound is
\begin{align}
\tau^* \leq \max \Bigg( \frac{1}{\mu \kappa \big( 1 + \frac{\gamma^2}{\kappa^2} \big)}, \frac{1}{2 \mu \kappa} \Bigg) .
\end{align}
However, as we saw in the previous section, both values are actually realized as the correlation time of a concrete observables, specifically $z_1 = x_1$ and $z_3 = x_1^2 + x_2^2$.
Since, by definition, the intrinsic correlation time is larger than the correlation time of any particular observable, we also have
\begin{align}
\tau^* \geq \max \big(\tau^{z_1},\tau^{z_3} \big) = \max \Bigg( \frac{1}{\mu \kappa \big( 1 + \frac{\gamma^2}{\kappa^2} \big)}, \frac{1}{2 \mu \kappa} \Bigg) .
\end{align}
Since $\tau^*$ is both upper and lower bounded by the expression on the right-hand side, we must have
\begin{align}
\tau^* = \max \Bigg( \frac{1}{\mu \kappa \big( 1 + \frac{\gamma^2}{\kappa^2} \big)}, \frac{1}{2 \mu \kappa} \Bigg) ,
\end{align}
which is the expression given in the main text.

\subsection{Estimate of the entropy production rate} \label{sec-parabolic-entropy}
As shown in \eqref{correlation-uncertainty-2} the main text, we can obtain a lower bound on the entropy production rate by considering the fluctuations of the time-average of an observable $z(\bm{x})$, as well as a bounded function $\chi_\text{min} \leq \chi(\bm{x}) \leq \chi_\text{max}$,
\begin{align}
\sigma_\text{st} \geq \frac{4}{\Delta \chi^2} \bigg( \frac{2 \text{Cov}_\text{st}(\chi,z)^2}{\text{Var}(\bar{z}_\tau)} - \av{\grad \chi \cdot \bm{B} \grad \chi}_\text{st} \bigg) \label{app-correlation-uncertainty}.
\end{align}
Here, we choose $z_1(x_1,x_2) = x_1$, and its truncation with range $\Delta$,
\begin{align}
\chi(x_1,x_2) = \left\lbrace \begin{array}{ll}
- \frac{\Delta}{2} &\text{for} \; x_1 < - \frac{\Delta}{2} \\[1ex]
x_1 &\text{for} \; - \frac{\Delta}{2} \leq x_1 \leq \frac{\Delta}{2} \\[1ex]
\frac{\Delta}{2} &\text{for} \; x_1 > \frac{\Delta}{2} .
\end{array} \right.
\end{align}
For this choice, we can compute \eqref{app-correlation-uncertainty} explicitly,
\begin{align}
\sigma_\text{st} \geq \hat{\sigma}(\delta)  = \frac{\kappa}{2 \delta^2} \Bigg( \bigg( 1 + \frac{\gamma^2}{\kappa^2} \bigg) \text{erf}(\delta)^2 - \text{erf}(\delta) \Bigg) \label{self-averaging-tur-parabolic},
\end{align}
where defined the dimensionless parameter $\delta = \sqrt{\kappa/(8T)} \Delta$ and $\text{erf}(y)$ denotes the error function.
The parameter $\delta$, which corresponds to the range of the truncation, can be chosen arbitrarily and we can thus maximize \eqref{self-averaging-tur-parabolic} with respect to $\delta$ and define $\hat{\sigma} = \max_\delta \hat{\sigma}(\delta)$.
The ratio of $\hat{\sigma}$ and the true value of the entropy production rate is show in Fig.~\ref{fig-parabolic-entropy} of the main text.

\subsection{Reduced transition probability} \label{sec-parabolic-path}
In the steady-state, the entropy production can be expressed in terms of the ratio of the probabilities of the forward and backward trajectory
\begin{align}
\Sigma = \tau \sigma_\text{st} = \int d\hat{\bm{x}} \ln \bigg( \frac{\mathbb{P}(\hat{\bm{x}})}{\mathbb{P}(\hat{\bm{x}}^\dagger)} \bigg) \mathbb{P}(\hat{\bm{x}}) \label{entropy-path} .
\end{align}
Here $\hat{\bm{x}} = (\bm{x}(t))_{t \in [0,\tau]}$ denotes the trajectory of the system during the observation interval and $\hat{\bm{x}}^\dagger = (\bm{x}(\tau-t))_{t \in [0,\tau]}$ the time-reversed trajectory.
Since the system is Markovian, this can be expressed in terms of the transition probability
\begin{align}
\sigma_\text{st} = \lim_{dt \rightarrow 0} \Bigg( \frac{1}{dt} \int d\bm{x} \int d\bm{y} \ \ln \bigg(\frac{p_{dt}(\bm{x} \vert \bm{y}) p_\text{st}(\bm{y})}{p_{dt}(\bm{y} \vert \bm{x}) p_\text{st}(\bm{x})} \bigg) p_{dt}(\bm{x} \vert \bm{y}) p_\text{st}(\bm{y}) \Bigg) .
\end{align}
In our case, since the system is two-dimensional, we can in principle calculate the entropy production rate by resolving the short-time behavior of both coordinates of the particle's position.
By contrast, the estimate \eqref{app-correlation-uncertainty} only relies on a measurement of one of the two coordinates, since both the observable and its truncation only depend on $x_1$.
In principle, we can also obtain such an estimate starting form \eqref{entropy-path}.
By integrating out the coordinate $x_2$ in the trajectory probability $\mathbb{P}(\hat{x}_1,\hat{x}_2)$, we obtain the reduced trajectory probability $\mathbb{P}(\hat{x}_1)$, which specifies the probability of observing a given trajectory for the $x_1$ coordinate of the particle.
From the information processing inequality, this reduced path probability provides a lower bound on the entropy production
\begin{align}
\Sigma \geq \int d\hat{x}_1 \ln \bigg( \frac{\mathbb{P}(\hat{x}_1)}{\mathbb{P}(\hat{x}_1^\dagger)} \bigg) \mathbb{P}(\hat{x}_1) \label{entropy-path-reduced} .
\end{align}
For the entropy production rate, the corresponding lower bound involves the marginal two-point probability density
\begin{align}
p_{dt}^{(1)}(x_1;y_1) = \int dx_2 \int dy_2 \ p_{dt}(x_1,x_2 \vert y_1,y_2) p_\text{st}(y_1,y_2) ,
\end{align}
in terms of which we have
\begin{align}
\sigma_\text{st} \geq \lim_{dt \rightarrow 0} \Bigg( \frac{1}{dt} \int dx_1 \int dy_1 \ \ln \bigg( \frac{p_{dt}^{(1)}(x_1 ; y_1)}{p_{dt}^{(1)}(y_1 ; x_1)} \bigg) p_{dt}^{(1)}(x_1 ; y_1) \Bigg) .
\end{align}
Using \eqref{steady-state-parabolic} and \eqref{transition-parabolic}, we can evaluate the marginal two-point density explicitly, the result is
\begin{align}
p_{t}^{(1)}(x_1;y_1) & = \frac{\kappa}{2 \pi T \sqrt{1 + e^{-2 \mu \kappa t} \big(\sin(\mu \gamma t)^2 - 1 \big)}} \label{reduced-transition} \\
& \hspace{2cm} \times \exp \Bigg[ -\frac{\kappa}{2 T \big( 1 + e^{-2\mu \kappa t} (\sin(\mu \gamma t)^2 - 1) \big)} \Big( x_1^2 + y_1^2 - 2 e^{-\mu \kappa t} x_1 y_1 \cos(\mu \gamma t) \Big) \Bigg] \n .
\end{align}
This is a Gaussian distribution with zero mean and covariance matrix
\begin{align}
\bm{\Xi} = \frac{T}{\kappa} \begin{pmatrix} 1 & e^{-\mu \kappa t} \cos(\mu \gamma t) \\ e^{-\mu \kappa t} \cos(\mu \gamma t) & 1 \end{pmatrix} .
\end{align}
While this depends on $\gamma$ and is different from the equilibrium result, it is symmetric under exchanging $x_1$ and $y_1$.
This implies that, when only observing the coordinate $x_1$, the trajectory probability is time-reversal symmetric, and the estimate on the entropy production obtained by using \eqref{entropy-path-reduced} vanishes.
This is reasonable considering the geometric structure of the driving force:
For a given $x_1$, there is an equal probability of finding respective positive or negative values of $x_2$, at which the driving force is equal but opposite, and thus the there is no net bias in the transition probability of $x_1$.
It seems surprising that, even though the trajectory probability of $x_1$ is time-reversal symmetric, the estimate \eqref{app-correlation-uncertainty}, which only depends on $x_1$ and can be obtained from the trajectory probability, yields a non-zero estimate on the entropy production rate.
The reason for this apparent contradiction is that the non-equilibrium nature of \eqref{reduced-transition} appears not only via its time-reversal symmetry, but also in the form of oscillations in the covariance matrix.
These oscillations are what gives rise to the accelerated self-averaging and the non-zero estimate on the entropy production via \eqref{app-correlation-uncertainty}.

\bibliography{bib}

\end{document}